\documentclass[twocolumn,superscriptaddress,amsmath,amssymb,aps,longbibliography,nofootinbib,prl,10pt]{revtex4-1}
\usepackage{txfonts}
\usepackage{graphicx}
\usepackage{bm}
\usepackage{booktabs}
\usepackage[dvipsnames]{xcolor}
\usepackage{braket}
\usepackage{dcolumn} 
\usepackage[colorlinks=true,urlcolor=blue,citecolor=blue,linkcolor=blue]{hyperref}

\begin{document}

\title{A low-symmetry ground state of dense two-dimensional hydrogen}

\author{Cesare Cozza}
\email{cesare.cozza@uzh.ch}
\affiliation{Department of Astrophysics, University of Zürich, Winterthurerstrasse 190, 8057 Zürich, Switzerland}
\affiliation{Scuola Internazionale Superiore di Studi Avanzati (SISSA), Via Bonomea 265, 34136 Trieste, Italy}

\author{Chris J. Pickard}
\affiliation{Department of Materials Science \& Metallurgy, University of Cambridge,
27 Charles Babbage Road, Cambridge CB3 0FS, United Kingdom}
\affiliation{Advanced Institute for Materials Research (WPI-AIMR), Tohoku University, Sendai, Miyagi, Japan}

\author{Guglielmo Mazzola}
\email{gmazzola@sissa.it}

\affiliation{Department of Astrophysics, University of Zürich, Winterthurerstrasse 190, 8057 Zürich, Switzerland}
\affiliation{Scuola Internazionale Superiore di Studi Avanzati (SISSA), Via Bonomea 265, 34136 Trieste, Italy}

\date{\today}

\begin{abstract}

Hydrogen under pressure is an extremely complex system featuring molecular dissociation, metallization, and anomalous melting.  While bulk hydrogen has been studied for nearly a century, its two-dimensional counterpart remains unexplored. Using first-principles structural searches and relaxations in supercells containing up to 512 atoms, we identify ground-state structures of dense two-dimensional hydrogen showing no evidence of long-range crystalline order over the simulated length scales. Within density functional theory, these structures have lower enthalpy than all crystalline candidates considered over an intermediate pressure interval between molecular and  atomic crystals. This preference already emerges with classical nuclei and persists as the supercell size increases, while the dominant structure-factor peaks grow substantially more slowly than in the crystalline reference phases. Although crystals with very large primitive cells cannot be rigorously excluded, these findings support the possibility of a disordered ground state and identify dense two-dimensional hydrogen as a promising setting for investigating competition between crystallization and structural disorder.
\end{abstract}

\maketitle

\textit{Introduction.} The dissociation and metallization of hydrogen under pressure has been a central problem in condensed-matter physics~\cite{RevModPhys.76.981} since the early prediction of Wigner and Huntington ~\cite{Wigner1935}. Despite extensive theoretical and experimental efforts, the high-pressure phase diagram in three dimensions remains only partially resolved, with ongoing debate on the nature of the molecular-to-atomic transition, the onset of metallic behavior~\cite{McMinis2015,Monserrat2018,Monacelli2023,DalladaySimpson2016}, and a possible low-temperature melting at high-pressures~\cite{Chen_2013,Liu2013,PhysRevB.111.104102,47cf-cj2s}.
Notably, decades ago, Ashcroft and coworkers suggested that the insulator-to-metal transition in hydrogen could proceed through an intermediate liquid phase,~\cite{Ashcroft2004,Babaev_2004} or non-crystalline or highly degenerate structures near dissociation ~\cite{Labet2012}. In three-dimensional  hydrogen (3DH), this possibility has been initially supported by both numerical and experimental observations of a reentrant melting line at higher temperature~\cite{Bonev2004, Deemyad2008, Zha2017, Cheng2020, Zong2020}. However, recent room-temperature experiments report no evidence of disordering upon compression at room-temperature~\cite{DalladaySimpson2016}, suggesting that the re-entrant melting line does not reach a zero temperature ground state~\cite{Cheng2020}.
 
Reduced dimensionality can qualitatively alter the balance between bonding, packing, and quantum fluctuations~\cite{KosterlitzThouless1972}. This motivates the exploration of hydrogen in two dimensions, looking for new phases of matter. Here we specifically concentrate on hydrogen monolayer: the nuclei are fixed to lay on a plane, but the electronic structure is solved in 3D.

Existing studies of two-dimensional hydrogen (2DH) are limited to a small number of non-relaxed high-symmetry candidate structures ~\cite{Biborski2017,Calcavecchia2017}, or to very-low pressure~\cite{Boninsegni2004,Cazorla2008}.  
This situation contrasts with recent
progress in other two-dimensional systems\cite{Li2021PhaseTransitions2D}, where phase diagrams have been
mapped with high accuracy, most notably for confined water
~\cite{Chen2016,Kapil2022,Lin2023,Jiang2021Freezing2DWater} and two-dimensional $^4$He
~\cite{Giorgini1996,Linteau2025}.
Given that even the  3DH phase diagram is still under debate, it is therefore important to determine whether reduced dimensionality stabilizes qualitatively distinct ground states or modifies the mechanism of metallization.

Here we provide a comprehensive \textit{ab initio} investigation of the zero-temperature phase diagram of 2DH, spanning the transition from molecular to atomic phases. We consider both candidate structures obtained from  random structure searching (AIRSS)~\cite{Pickard2006,Pickard2011} and supercell configurations inspired by candidate 3D crystals~\cite{Pickard2007} (see below). Remarkably, we find that  disordered supercell structures become enthalpically favoured compared to crystalline phases at the onset of molecular dissociation, suggesting a qualitatively different route to metallization in reduced dimensionality.
Nuclear quantum effects further stabilizes this  phase.

\begin{figure*}[ht!]
    \centering
    \includegraphics[width=1.0\textwidth]{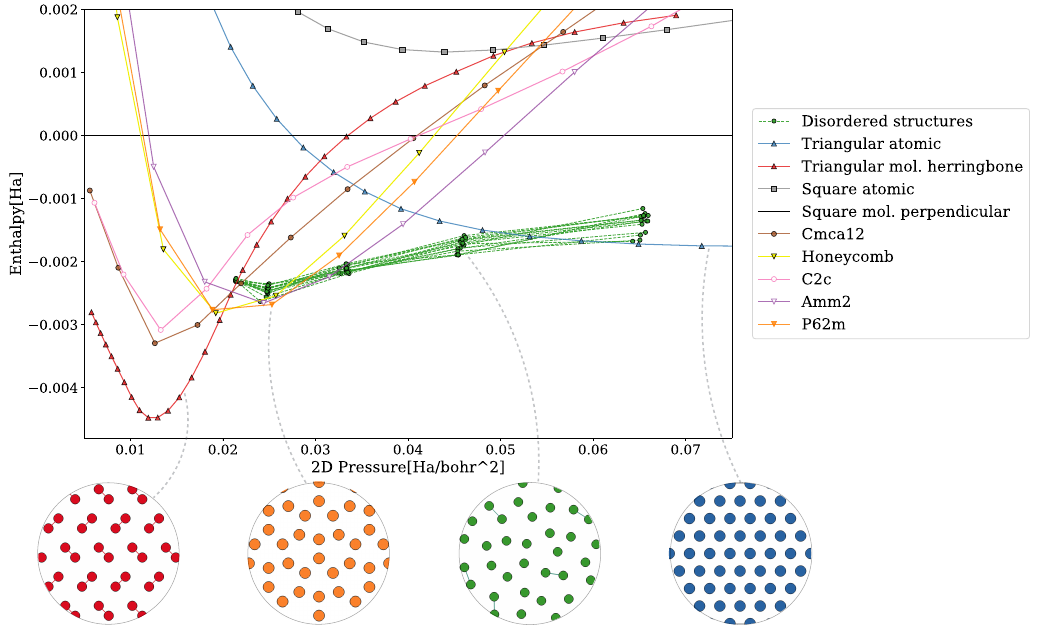}
    \caption{\textbf{Phase diagram of 2D Hydrogen at zero temperature.} \emph{(Top:)} Enthalpy per atom vs pressure of several candidate structures.  The enthalpies are calculated as difference compared with a common reference: a square molecular perpendicular lattice (see Supplementary Material).
    We observe that four phases become enthalpically favored upon compression. In order of increasing pressure, these are: molecular herringbone (red), $P62m$ (orange), disordered structures (green), triangular atomic (blue).
    This particular phase diagram has been constructed using 288 atoms supercells. The green dashed lines connecting different instances (obtained from different random initial configurations) of relaxed disordered structures are simply a guide to the eye.
    \emph{(Bottom:)} Images of the ground state structures. }
    \label{fig:phase_diagram}
\end{figure*}

\textit{Methods.} We restrict our analysis to two-dimensional crystalline arrangements of the protons, while solving the electronic structure in three dimensions within plane-wave density functional theory (DFT) at the PBE level ~\cite{Perdew1996}.
In our setup, we employ three-dimensional simulation cells in which each structure is initialized and constrained to move within the 
$x-y$ plane, while a large interplanar spacing is maintained along the 
$z$ direction.
Simulations are performed using supercells containing up to 512 atoms with Brillouin-zone sampling on a $6 \times 6 \times 1$ $k$-point grid (see Supplementary Material ~\cite{supmat} for details). Finite-size effects are assessed by varying the system size $N$ from 48 to 512 atoms.
DFT simulations are performed with {\sc Quantum ESPRESSO} software package (\textsf{pw.x} code) ~\cite{Giannozzi_2017} using PAW pseudopotential ~\cite{BlochlPAW1004} from pslibrary ~\cite{DalCorso2014_psl} and energy and density cutoff respectively of 140.0 and 1120.0 Ry. 
We further cross-check our results with CASTEP ~\cite{Clark2005CASTEP}.

The main result is already obtained by treating the nuclei as classical particles. Nuclear quantum effects are later assessed via path-integral molecular dynamics (PIMD) simulations, which we use  to test the dynamical stability of crystalline phases at pressures where disordered structures are  enthalpically competing.
For each structure, we compute both the internal energy $E$ and the enthalpy $H = E + pA$, where $A$ is the area of the cell, $p$ is the two-dimensional pressure (see Supplementary Material ~\cite{supmat} for its definition).  
This distinction is essential, as our ranking depends on whether we evaluate energy or enthalpy; the latter being 
the correct thermodynamic quantity for determining phase stability under a constant applied pressure. The results concerning $E$ are provided in the Supplementary Material~\cite{supmat}.

At each density, structural relaxations are performed starting from a broad set of initial configurations. We consider: (i) molecular crystals, where $\mathrm{H}_2$ units are arranged on square, triangular, and hexagonal lattices with different orientations (parallel, orthogonal, staggered, and random); (ii) atomic crystals, including square, triangular, honeycomb, and disordered configurations;  (iii) structures derived from three-dimensional phases: structures such as \emph{Cmca}-12, \emph{C2/c} are made of (almost) planary 2D lattices stacked along the z directions; and (iv) AIRSS-predicted candidates at high pressure (e.g., $P62m$, $Amm2$, $P4/mmm$, $Pbam$, $Cmmm$).
Notably, AIRSS also returns the structures that have been ``manually'' proposed, at the low and the high ends of the pressure domain. 
For disordered configurations, relaxations are initialized from multiple random seeds (typically 16), and statistical uncertainties are reported accordingly.
As detailed in the Supplementary Material, computing the thermodynamic properties of the crystalline phases using supercell setup yields identical results to evaluating the corresponding primitive cells with an appropriately scaled k-point mesh.

\textit{Results.} We perform structural relaxations and observe that some configurations undergo only minor local rearrangements, whereas others exhibit more significant changes.
We inspect the relaxed structures and
compute the enthalpy--pressure curves for all resulting phases (see Fig.~\ref{fig:phase_diagram})

We find that only \text{four} phases are enthalpically favored over the considered pressure range. At low pressure, the ground-state phase is \emph{molecular}: the centers of mass of the molecules lie on the sites of a triangular lattice, and the molecular orientations are parallel within each (horizontal) line that lies parallel to the base of the triangles (we refer to this arrangement as a ``herringbone'' lattice, see Fig.~\ref{fig:phase_diagram}). This phase is stable up to $0.02$~Ha/Bohr$^2$.
At higher pressure, a structure with in-plane symmetry consistent with the $P62m$ symmetry group becomes stable up to approximately $0.03$~Ha/Bohr$^2$. This \emph{atomic} structure corresponds to a distorted honeycomb lattice, similar to a plane exfoliated from the $C2/c$ structure, which has been proposed as a candidate for phase~III of 3D hydrogen~\cite{Pickard2007,Monacelli2023}.
The $Amm2$ structure is enthalpically compatible with $P62m$ close to the transition. 
Both $P62m$ and $Amm2$ have been  discovered by AIRSS.

Structures without any long-range order within the supercell become favored in the pressure range between $0.029(1)$ and $0.054(4)$~Ha/Bohr$^2$, which correspond to a density of about 1.17 - 0. Bohr in units of Wigner-Seitz radius ($r_s = \sqrt{A/(\pi N)}$ in 2D). This is the main finding reported in this manuscript. 
These structures have been found by relaxing from 16 different atomic configurations where atoms were uniformly sampled in the box, for each investigated system size.  In this range, we observe no indication of crystallization (even considering defects). Most of the resulting structures show lower enthalpy than the crystalline ones. To ensure that no competitive crystalline phases were overlooked, extensive AIRSS searches ($\sim 10^4$ candidate structures) were targeted specifically within this pressure window (detailed in the Supplementary Materials \cite{supmat}).

Interestingly, we observe spontaneous crystallization (from completely disordered initial structures) into a perfect herringbone phase below $0.02$~Ha/Bohr$^2$, and a defective atomic triangular above $0.07$~Ha/Bohr$^2$.

Notice how several crystals (honeycomb, $P62m$, and $Amm2$) become enthalpically degenerate at the onset of the disordered phase, while, at about $0.04$~Ha/Bohr$^2$, the disordered structures show their maximum enthalpy difference compared to the competing crystalline phases.

Finally, above $0.054(4)$~Ha/Bohr$^2$, a clearly crystalline phase again becomes the ground state, adopting a simple triangular lattice.
In Fig.~\ref{fig:phase_diagram} we present the equations of state for all low-enthalpy structures, including some of those that are not ground states. For completeness, more candidate's equation of states are provided in the Supplementary Materials. 

Microscopically, the low-symmetry intermediate seems to bridge the transition from crystals with coordination number 3 (honeycomb, $P62m$ and $Amm2$) to 6 (triangular).

A preliminary characterization of the insulator-to-metal transition is obtained by calculating the density of states at the Fermi level within the PBE level of theory (See Supplementary Material).
Interestingly, according to this metric, metallization occurs at the transition from the molecular phase to the first atomic structure, $P62m$. Therefore, within this mean-field description, the low-symmetry structures are also predicted to be metallic.

\begin{figure}[ht!]
    \centering
\includegraphics[width=0.95\columnwidth]{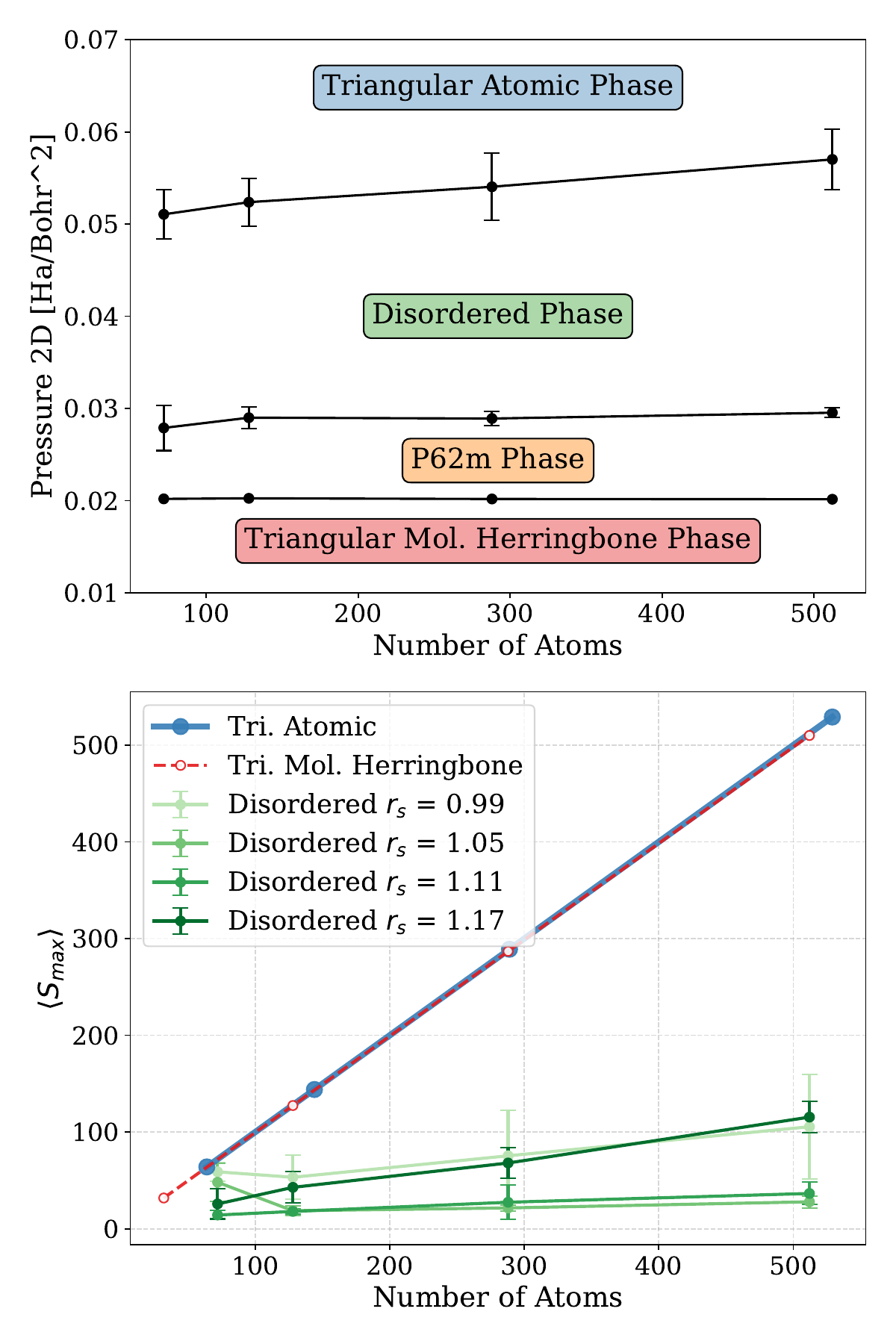}
    \caption{\textbf{Size scalings.} \emph{(Top:)} Phase boundaries as a function of the system's size $N$.  
    We consider here only supercells from about 70 atoms to 512 atoms, since they provide consistent phase diagrams. The phase diagram obtained with a smaller set-up is qualitatively different.
    \emph{(Bottom:)} Maximum of Structure factor $S_{max}$ as a function of $N$ for the most stable lattices across the phase diagram. The disordered structures correspond to different densities, characterized by the Wigner--Seitz radius $r_s$ (in Bohr). The $r_s = 0.99$ Bohr and $r_s = 1.17$ Bohr datasets correspond to the disordered structures around $p_{2D} = 0.25$ and $0.065$ Ha/Bohr$^2$, respectively (see Fig.~\ref{fig:phase_diagram}). The $r_s = 0.99$ and $r_s=1.17$ sets, are close to the stability limits of the crystalline phases and contain structures that relax towards defective crystals. This accounts for the residual increase of $S_{\mathrm{max}}$ with $N$.
        }
    \label{fig:scaling}
\end{figure}

\textit{Finite size effects.} We perform a finite-size scaling analysis to assess the stability of the  phases toward the thermodynamic limit. 
To do so, we compute the phase diagram using different supercell sizes, from 48 to 512 atoms.
Above $\sim$70 atoms the  phase diagrams and boundaries remain stable with increasing system size, indicating the stabilization of a disordered phase is not artifact of the finite simulation cell (see Fig.~\ref{fig:scaling}). We also note that, in two dimensions, the thermodynamic limit is reached more rapidly than in three dimensions.
Crucially, the phase diagram computed using 48-atom supercells is qualitatively different from those obtained with larger supercells. Some new crystalline structures emerge as ground states ($Cmca12$, $Amm2$), and, most importantly, the low-symmetry phase is favored only within a very narrow pressure window (see Supplementary Materials). This finding highlights the importance of overcoming finite-size effects through the use of large supercells. 

Further, we compute the maximum structure factor contribution $S_\textrm{max}$, defined in the Supplementary Materials~\cite{supmat}, as a function of $N$. We find that $S_\textrm{max}$ increases as $N$ as expected for the crystalline phases, while it grows much more slowly for the samples of the disordered phase. This provides a more rigorous indication of absence of long-range order than a simple visual inspection of the configurations, and does not reveal signatures of a quasi-crystalline phase \cite{PhysRevLett.53.2477}. Nevertheless, confirming the absolute absence of any crystalline ground state is computationally bounded; we cannot categorically exclude that searches utilizing larger atom counts might find a stable crystal with a very large primitive cell.

\textit{Nuclear quantum effects (NQE).}
So far, we have treated the nuclei as classical particles, which, in a realistic description of a material at zero temperature, can be regarded only as an approximation.
Following the scaling argument of Ashcroft and coworkers~\cite{Labet2012}, one may expect the proton zero-point energy to be reduced in a disordered structure relative to a crystalline phase,  enthalpically favoring even more a disordered (or possibly liquid) ground state. At the same time, NQE may also act in the opposite direction by promoting bond symmetrization, thereby possibly stabilizing structures with higher symmetry~\cite{Benoit1998Tunnelling,Errea_2016}.
We test the stability of the closest crystalline candidate upon inclusion of proton quantum effects. To this end, we perform path-integral simulations with 120(128) atoms cells for a $P62m$(disordered) initial structure, at a density for which $P62m$ is the classical ground-state structure and observe its rapid melting into a disordered phase. In the Supplementary Material, we report the proof of rapid decay and equilibration of the structure factor, as well as the radial and angular pair-distribution functions.

\textit{Discussions.}
We report a dense two-dimensional phase of hydrogen in which
disordered proton configurations are already favored at the
Born--Oppenheimer, classical-nuclei level. 
We first note that this is not in contradiction with existing mathematical
results on crystallization, which have only been established for 
classes of pairwise interaction potentials~\cite{Theil2006,BlancLewin2015}, whereas an \textit{ab initio}
energy surface is intrinsically many-body.
From a computational perspective, the new low-symmetry phase appears to be stable only in supercells containing more than $\sim 50$ atoms. 
Obviously, we cannot rule out the possibility of a crystalline structure with a primitive cell larger than our simulation supercell.
Dense alkali metals demonstrate that very large cell and incommensurate
elemental structures are possible, and therefore no finite-cell calculation
can categorically exclude them
\cite{Gregoryanz2008StructuralDiversity,Lundegaard2009IncommensurateNa,
McMahon2006KIII}. However, the largest such examples occur in three
dimensions and, for sodium, close to a finite-temperature melting minimum
\cite{Gregoryanz2008StructuralDiversity}.
At fixed atom number, the present two-dimensional calculations cover a
much larger linear range. Moreover, coreless hydrogen lacks the ionic-core
overlap and associated interstitial-electron localization implicated in the
structural complexity of compressed alkali metals
\cite{NeatonAshcroft2001Sodium,Marques2011DenseLithium}.
Larger supercells calculations could be achieved with
 machine-learning potentials~\cite{BehlerCsanyi2021,Kocer2022}.

The disordered phase of 2DH is different from
 canonical zero-temperature quantum liquids, $^{3}$He and $^{4}$He, whose
stability is conventionally attributed to large
zero-point motion overcoming weak van der Waals binding~\cite{Kent1993}.
It also separates the present mechanism from ordinary low-melting elemental
liquids such as Hg,~\cite{KozinHansen2013}, and  Li~\cite{Guillaume2011}, where the liquid phase still terminates
at finite temperature. 
At the same time, proton-disordered configurations of ordinary water ice are not expected to constitute the thermodynamic ground state at zero temperature. At ambient pressure, proton-disordered ice Ih undergoes proton ordering toward ice XI, which is identified as the ground-state structure\cite{Singer2005,Gasser2021}. The persistence of proton disorder at low temperature is therefore generally associated with kinetic arrest rather than with an intrinsically disordered ground state.

Further,
molecular para-hydrogen in two-dimensions
 predicted to crystallize at $T=0$, with no metastable liquid
phase, and the one-dimensional equilibrium phase is likewise crystalline
~\cite{Boninsegni2004,Boninsegni2013}. 
Finally, there is no numerical evidence for ground state liquid in three-dimensional hydrogen, 
 also at ultra-high pressures, deep into the metallic regime, as recently found~\cite{47cf-cj2s}.
Even if 3DH eventually becomes a ground state metallic fluid, the 2D disordered phase we report is different as it appears as \emph{intermediate} between the molecular and the atomic phases.

The implication is that dense
two-dimensional hydrogen realizes a distinct route to structural disorder.
This establishes
dimensional confinement as a control parameter for suppressing crystallization
in hydrogen and suggests routes to
engineer low-temperature liquids~\cite{Liu2025}.

Proton quantum fluctuations  further stabilize the phase, rather than being solely responsible for melting, although we leave for future work a comprehensive characterization of the phase diagram including quantum proton effects, and a precise determination of a liquid phase.

Our conclusions are based on DFT-PBE energetics. 
A next step is to
use more
accurate electronic-structure methods, such as quantum Monte Carlo (QMC)~\cite{nakano2020turborvb,cozza_denser_2026,linteau2026neural} or
exchange--correlation functionals recently known to perform better for dense hydrogen~\cite{cozza_denser_2026}.
However, it would be surprising that a small quantitative change in the potential-energy surface could deliver a
qualitative change of this picture, as this is not occurring in three-dimensional hydrogen both in the solid~\cite{Monacelli2023}, and liquid phases~\cite{Knudson2015,mazzola2018}.
Moreover, the significance of our finding is primarily qualitative rather than
quantitative. Since free-standing 2DH cannot be realized experimentally, any
experimental implementation will necessarily involve a specific substrate, which
should ultimately be included explicitly in the simulations. The confinement surfaces are
likely to modify the energetics more strongly than, for example, the residual
difference between QMC and DFT for an idealized free-standing layer.
Moreover, beyond providing a realistic description of hydrogen, this result suggests that there exists at least one $n$-body potential capable of giving rise to disorder.

On a more fundamental level, it will be important to identify the microscopic
mechanism that stabilizes such an uncommon phase of matter and, in particular,
to pinpoint the role played by reduced dimensionality, and explore the realistic regimes beyond the single atomic layer~\cite{Li2021PhaseTransitions2D}.

\begin{acknowledgments}
\textit{Acknowledgments.} We acknowledge useful discussions with Michele Ceriotti,  Marcello Porta, Giuseppe Carleo, Stefano de Gironcoli, Sandro Scandolo, Alessandro Laio, and Stefano Baroni.
We gratefully acknowledge Ravit Helled for instrumental  support throughout this work.
C.C, G.M. acknowledge financial support from the Swiss National Science Foundation (Grant No. PCEFP2\_203455) and computational support from CSCS and Science Cluster at University of Z\"urich.
All structures found in this study are available at the following repository~\cite{github_cesare}.
\end{acknowledgments}

\bibliography{biblio_short}

\clearpage
\pagebreak
\widetext
\begin{center}
\textbf{\large Supplementary Material}
\end{center}
\setcounter{equation}{0}
\setcounter{figure}{0}
\setcounter{table}{0}
\setcounter{page}{1}
\renewcommand{\theequation}{S\arabic{equation}}
\renewcommand{\thefigure}{S\arabic{figure}}
\renewcommand{\bibnumfmt}[1]{[S#1]}

\section{Definition of the 2D Pressure}

In three dimensions, the pressure can be defined from the derivative of the internal energy with respect to a homogeneous volumetric strain as
\begin{equation}
P_{\mathrm{3D}} = \frac{1}{V}\frac{\partial E_{\mathrm{int}}}{\partial \epsilon_{\mathrm{3D}}},
\qquad [\mathrm{Ha/Bohr}^3],
\label{eq:p3d_num}
\end{equation}
where $\epsilon_{\mathrm{3D}} = \Delta V / V$ denotes the relative volume change.

By analogy, in two dimensions we define the 2D pressure as
\begin{equation}
P_{\mathrm{2D}} = \frac{1}{A}\frac{\partial E_{\mathrm{int}}}{\partial \epsilon_{\mathrm{2D}}},
\qquad [\mathrm{Ha/Bohr}^2],
\label{eq:p2d_num}
\end{equation}
where $\epsilon_{\mathrm{2D}} = \Delta A / A$ is the relative change in the in-plane area.

In practice, Eq.~\eqref{eq:p2d_num} can be evaluated numerically by applying a set of small homogeneous strains (typically seven points around the equilibrium configuration) and fitting the resulting energy–strain to obtain the derivative. This procedure increases the computational cost by a factor corresponding to the number of strain points, compared to evaluating the stress tensor directly via the Hellmann-Feynman theorem.

Within density functional theory, the 3D pressure is more conveniently obtained from the stress tensor as
\begin{equation}
P_{\mathrm{3D}} = -\frac{1}{3V}\left(\sigma_{xx} + \sigma_{yy} + \sigma_{zz}\right),
\label{eq:p3d_analytical}
\end{equation}
which can be generalized to two dimensions as
\begin{equation}
P_{\mathrm{2D}} = -\frac{1}{2A}\left(\sigma_{xx} + \sigma_{yy}\right).
\label{eq:p2d_analytical}
\end{equation}

\begin{figure}[h]
\centering
\includegraphics[width=0.5\linewidth]{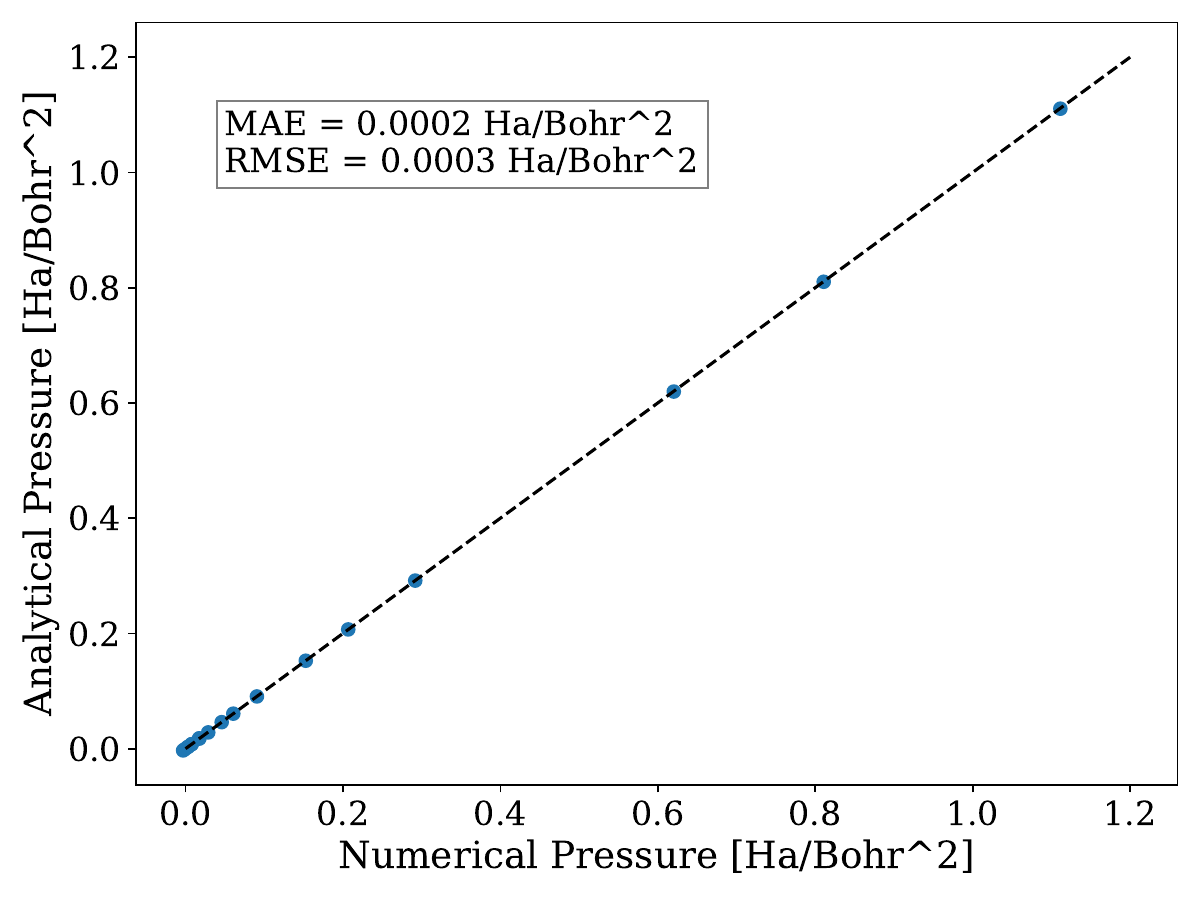}
\caption{Comparison between the 2D pressure computed numerically from finite strains [Eq.~\eqref{eq:p2d_num}] and analytically from the stress tensor [Eq.~\eqref{eq:p2d_analytical}]. Each point corresponds to a different atomic or molecular configuration.}
\label{fig:p2d}
\end{figure}

We have verified that the numerical definition in Eq.~\eqref{eq:p2d_num} and the analytical stress-tensor expression in Eq.~\eqref{eq:p2d_analytical} yield consistent results across a representative set of atomic and molecular configurations, as shown in Fig.~\ref{fig:p2d}.

\section{DFT Setup}
All density functional theory (DFT) calculations were performed using QuantumESPRESSO~\cite{Giannozzi_2017} (version 7.0, code \textsf{pw.x}) with a plane-wave basis set. We employ the PBE exchange-correlation functional~\cite{Perdew1996} together with projector augmented-wave (PAW) pseudopotentials~\cite{BlochlPAW1004} from the PSLibrary \cite{DalCorso2014_psl}, specifically \texttt{H.pbe-kjpaw\_psl.1.0.0.UPF}.

The kinetic energy cutoff for the plane-wave basis was set to 140 Ry, while the charge density cutoff was set to 1120 Ry. A Gaussian smearing of 0.002 Ry was used. Brillouin zone integrations were performed using a $6\times 6\times 1$ Monkhorst--Pack \textit{k}-point grid. To suppress spurious interactions between periodic images in the out-of-plane direction, a vacuum spacing of 24.0 Bohr was introduced along the cell height ($z$ direction).

\begin{figure}
\centering
\includegraphics[width=0.8\linewidth]{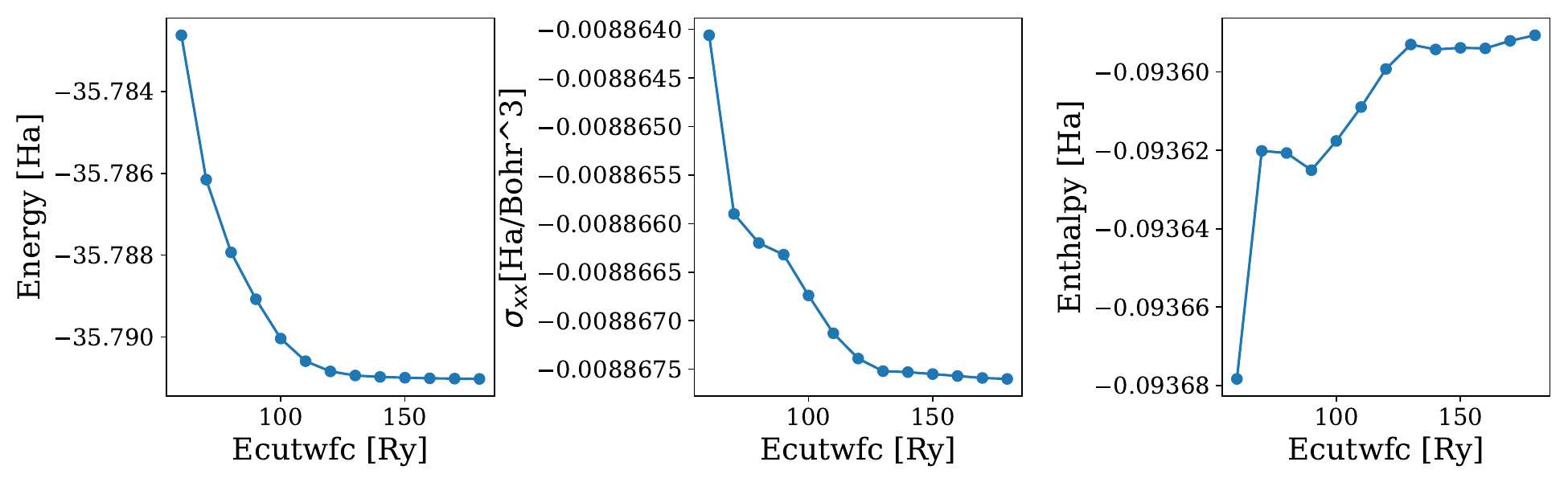}
\caption{Plane-wave energy cutoff convergence test for the total energy, stress tensor component $\sigma_{xx}$ (identical to $\sigma_{yy}$), and enthalpy. The chosen cutoff is 140 Ry.}
\label{fig:ecut_conv}
\end{figure}

\begin{figure}
\centering
\includegraphics[width=0.8\linewidth]{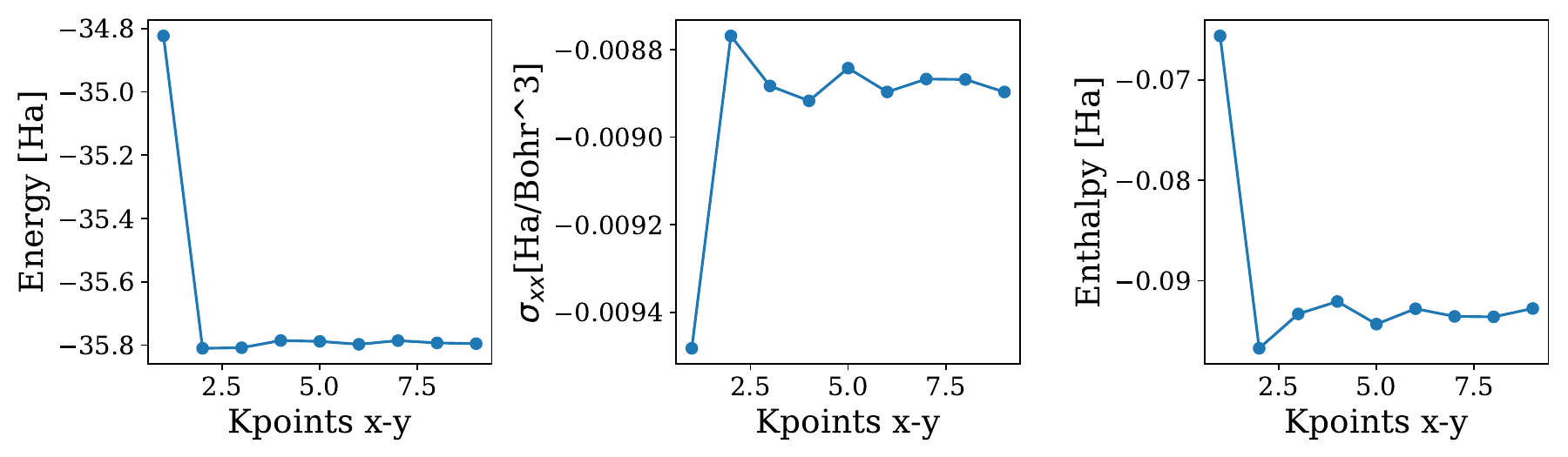}
\caption{\textit{k}-point convergence test for the total energy, stress tensor component $\sigma_{xx}$ (identical to $\sigma_{yy}$), and enthalpy. A $6\times 6\times 1$ grid was used in production calculations.}
\label{fig:kpts_conv}
\end{figure}

\begin{figure}
\centering
\includegraphics[width=0.8\linewidth]{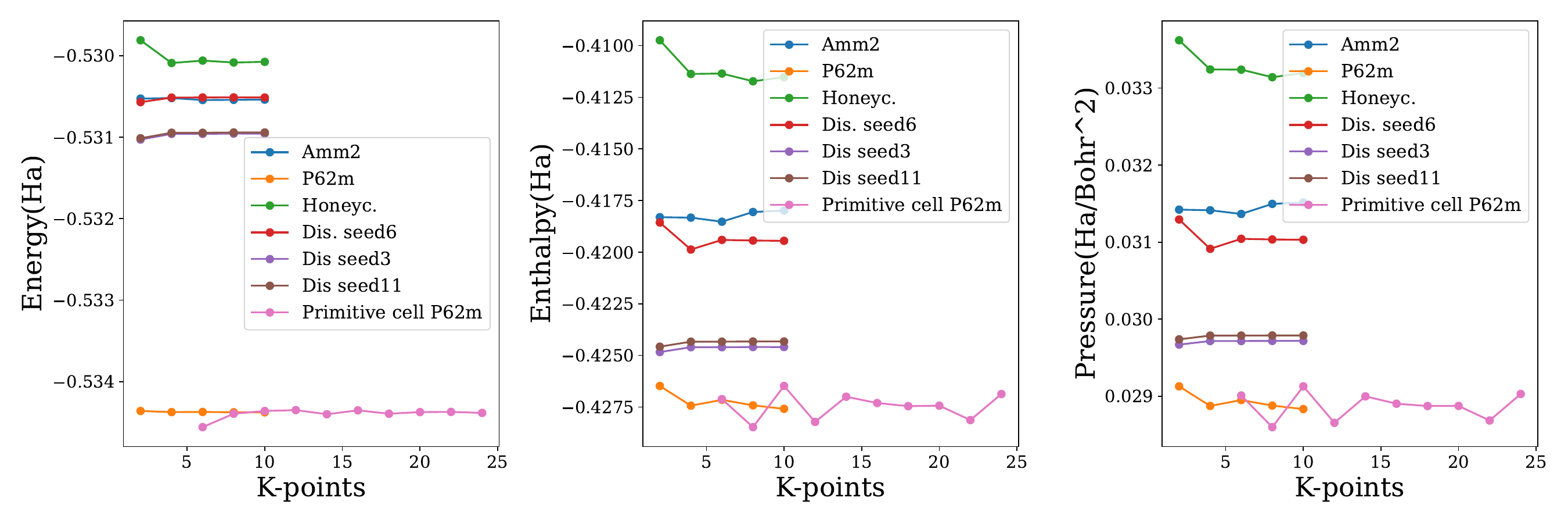}
\caption{Ordering of different lattices in a range of pressures around 0.030 $\text{Ha}/\text{Bohr}^2$ with varying k-points grid. In all cases, except for the pink data serie, we use supercells as reported in the main text. We test three crystalline and three randomly chosen non crystalline structures (labelled with their seed numbers). For $P62m$ we also use primitive cell with 6-atoms, expanding the k-points grid. This shows that, for crystalline lattices, using supercell with less k-points is equivalent to using primitive cells with more k-points above a certain threshold. }
\label{fig:ordering_vs_kpts}
\end{figure}

\begin{figure}
\centering
\includegraphics[width=0.6\linewidth]{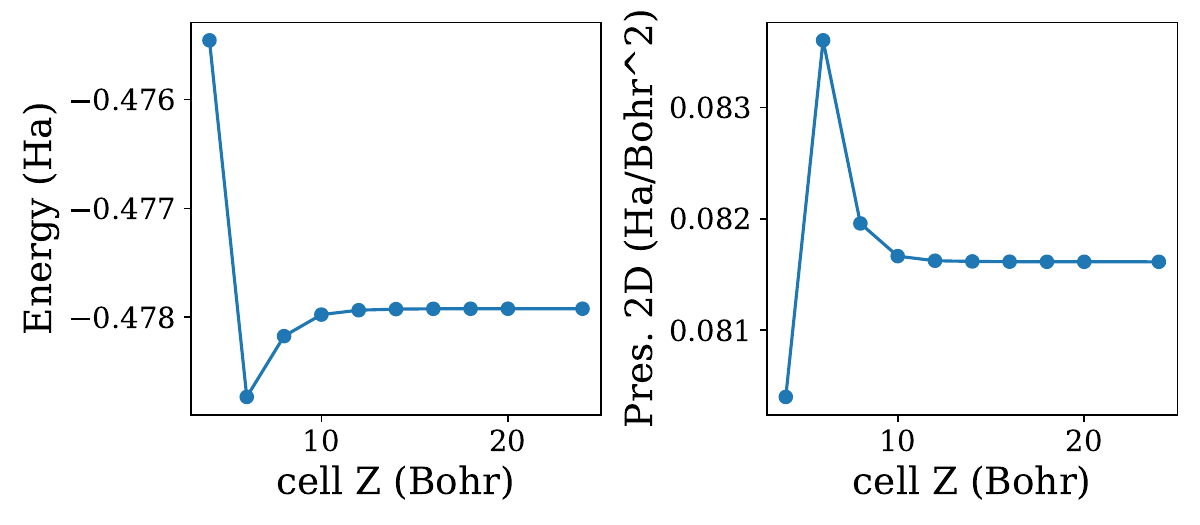}
\caption{Vacuum spacing convergence test with respect to total energy and 2D pressure. A vacuum height of 24.0 Bohr was chosen for production calculations.}
\label{fig:cellz_conv}
\end{figure}

Convergence with respect to the plane-wave cutoff, \textit{k}-point sampling, and vacuum spacing was carefully verified. In all cases, the charge density cutoff was fixed to eight times the kinetic energy cutoff. The corresponding convergence tests for the total energy, stress tensor components (with identical behavior for the $xx$ and $yy$ components), and enthalpy are shown in Figs.~\ref{fig:ecut_conv}–\ref{fig:cellz_conv}. \\
A further check is provided in Fig.\ref{fig:ordering_vs_kpts}, which demonstrates that the density of the k-point sampling does not affect the ordering of the various lattices in the phase diagram. With the 2D pressure fixed around 0.030 $\text{Ha}/\text{Bohr}^2$, the most promising lattice candidates were evaluated across an increasing k-point grid to confirm that their relative stability remains unchanged. These calculations primarily utilized a 128/120-atom supercells; however, we also verified that using the 6-atom primitive cell of one of the crystalline candidates (specifically, the $P62m$ structure) yields a 2D enthalpy in perfect agreement when the appropriate k-point sampling is applied.

\begin{figure}
\centering
\includegraphics[width=0.9\linewidth]{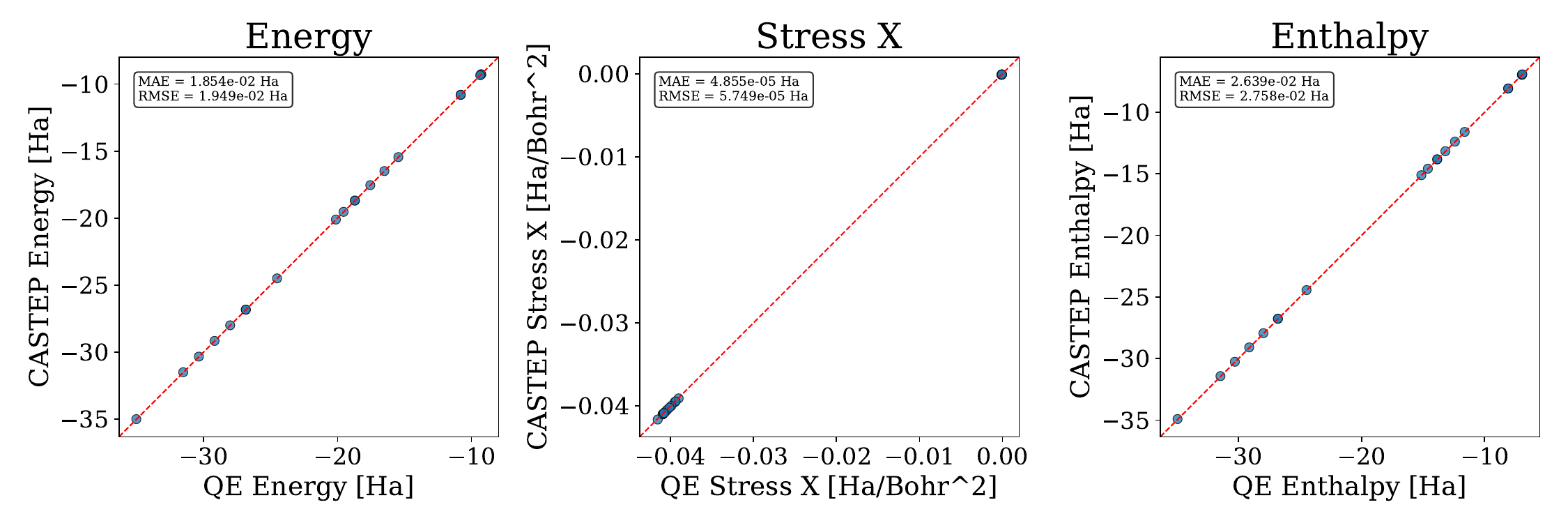}
\caption{Comparison of Energy, xx component of the stress tensor and enthalpy over a set of configurations generated through AIRSS search between QuantumESPRESSO and CASTEP.}
\label{fig:qe_vs_castep}
\end{figure}

To demonstrate the robustness of our previously defined setup, we verified the agreement between QuantumESPRESSO and CASTEP~\cite{castep} (version 26.11). For the CASTEP calculations, identical DFT parameters were used, with the exception of the pseudopotential, since PAW pseudopotentials are not available for this code. Instead, we used the ultrasoft pseudopotential \texttt{H.pbe-rrkjus\_psl.1.0.0.UPF} from pslibrary~\cite{DalCorso2014_psl}. As shown in Fig.~\ref{fig:qe_vs_castep}, we present a comparison between the two DFT codes, specifically evaluating energy, the xx component of the stress tensor, and enthalpy, across a set of configurations derived from an AIRSS search.

\clearpage
\section{Details of the Structure Factor Calculation}

To characterize translational order in the candidate structures, we compute the static structure factor in reciprocal space. The structure factor describes the intensity of radiation scattered by the system and is defined as

\begin{equation}
S(\mathbf{k}) = \frac{1}{N}
\left|
\sum_{j=1}^{N}
e^{i\mathbf{k}\cdot\mathbf{R}_j}
\right|^2 ,
\label{eq:Sq}
\end{equation}

where (N) is the number of atoms and ($\mathbf{R}_j$) denotes the position of atom (j).

The structure factor corresponds to the diffraction pattern that would be observed in a scattering experiment. In an ordered phase, the diffraction pattern exhibits Bragg peaks arising from constructive interference between periodically arranged atomic planes. The presence of sharp peaks at nonzero reciprocal lattice vectors is a hallmark of long-range crystalline order. By contrast, a disordered phase displays diffuse scattering with no pronounced peaks apart from the trivial forward-scattering contribution at the $(\Gamma)$ point (($\mathbf{k}=0$)).

\begin{figure}
\centering
\includegraphics[width=1.0\linewidth]{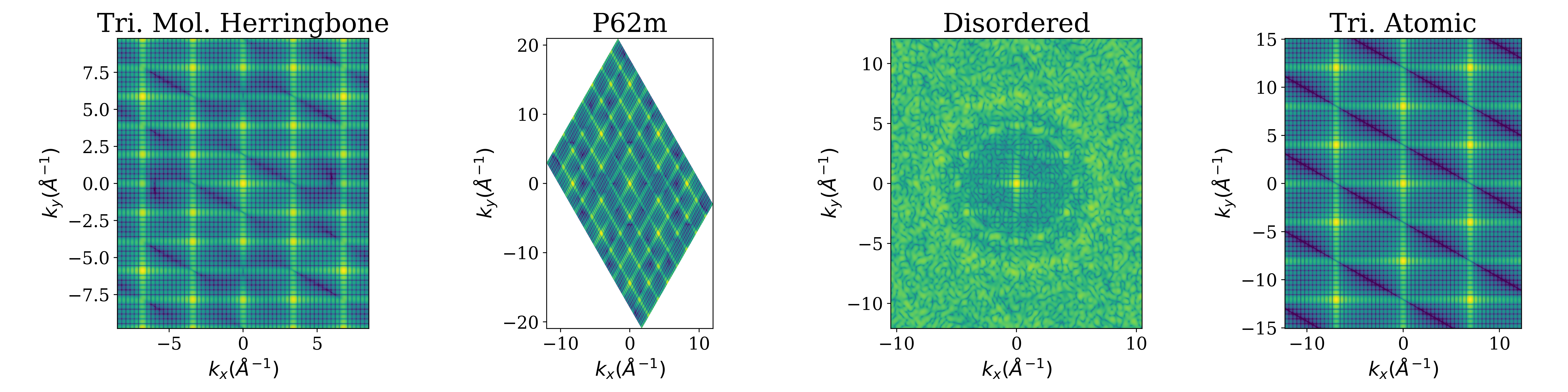}
\caption{Structure factors of the four favored lattice structures in the different pressure ranges.}
\label{fig:Structure_factor}
\end{figure}

The peak at $(\Gamma)$ is always present and has height $(S(\Gamma)=N)$, reflecting coherent forward scattering from all atoms in the system. As shown in Fig.~\ref{fig:Structure_factor}, the ordered structures: Triangular Atomic, $P62m$, and Triangular Molecular Herringbone, exhibit well-defined Bragg peaks, whereas the disordered structure displays diffuse scattering with no sharp peaks away from $(\Gamma)$.

Because the calculations are performed on finite supercells, the Bragg peaks remain finite rather than diverging as in the thermodynamic limit. Their height scales approximately linearly with the number of atoms. Finite-size effects are also responsible for the line-like features visible in the structure-factor maps.

We further employ the structure factor to perform a finite-size scaling analysis. By monitoring the maximum value of the structure factor, $S(\mathbf{k}_{max})$, as a function of system size, we can determine whether long-range order persists in the thermodynamic limit. Here, $\mathbf{k}_{max}$ denotes the reciprocal-space point corresponding to the largest nontrivial peak in the structure factor.

In an ordered phase, the height of the dominant Bragg peak scales linearly with the number of atoms,

\begin{equation}
S(\mathbf{k}_{max}) \sim N,
\end{equation}

whereas in a disordered liquid phase it approaches a constant value independent of system size,

\begin{equation}
S(\mathbf{k}_{\mathrm{max}}) \sim \mathrm{const}.
\end{equation}

Therefore, the scaling behavior of $S(\mathbf{k}_{max})$ provides a direct criterion to distinguish between ordered and disordered phases.

\section{Details of AIRSS}

We performed an \emph{ab initio} random structure search (AIRSS) \cite{Pickard2011} using \textsf{Quantum ESPRESSO} to systematically explore low-enthalpy 2D hydrogen structures over the pressure range of interest. All DFT parameters were kept identical to those described in the previous sections, except for the Brillouin zone sampling. In this case, the \textit{k}-point mesh was chosen according to a reciprocal-space density of 0.07 \AA$^{-1}$ in the $xy$ plane, while a single \textit{k}-point was used along the out-of-plane ($z$) direction.

At low and high pressures, AIRSS correctly recovers the triangular molecular herringbone phase and the triangular atomic phase, respectively, in agreement with our previous results. In the intermediate-pressure regime, where disordered structures are competitive, we selected four representative pressures and performed additional AIRSS searches. These calculations were carried out both with and without enforcing atomic or molecular symmetry constraints, and both using or not the relax-and-shake (RASH) algorithm.

\begin{figure}
    \centering
    \includegraphics[width=0.49\linewidth]{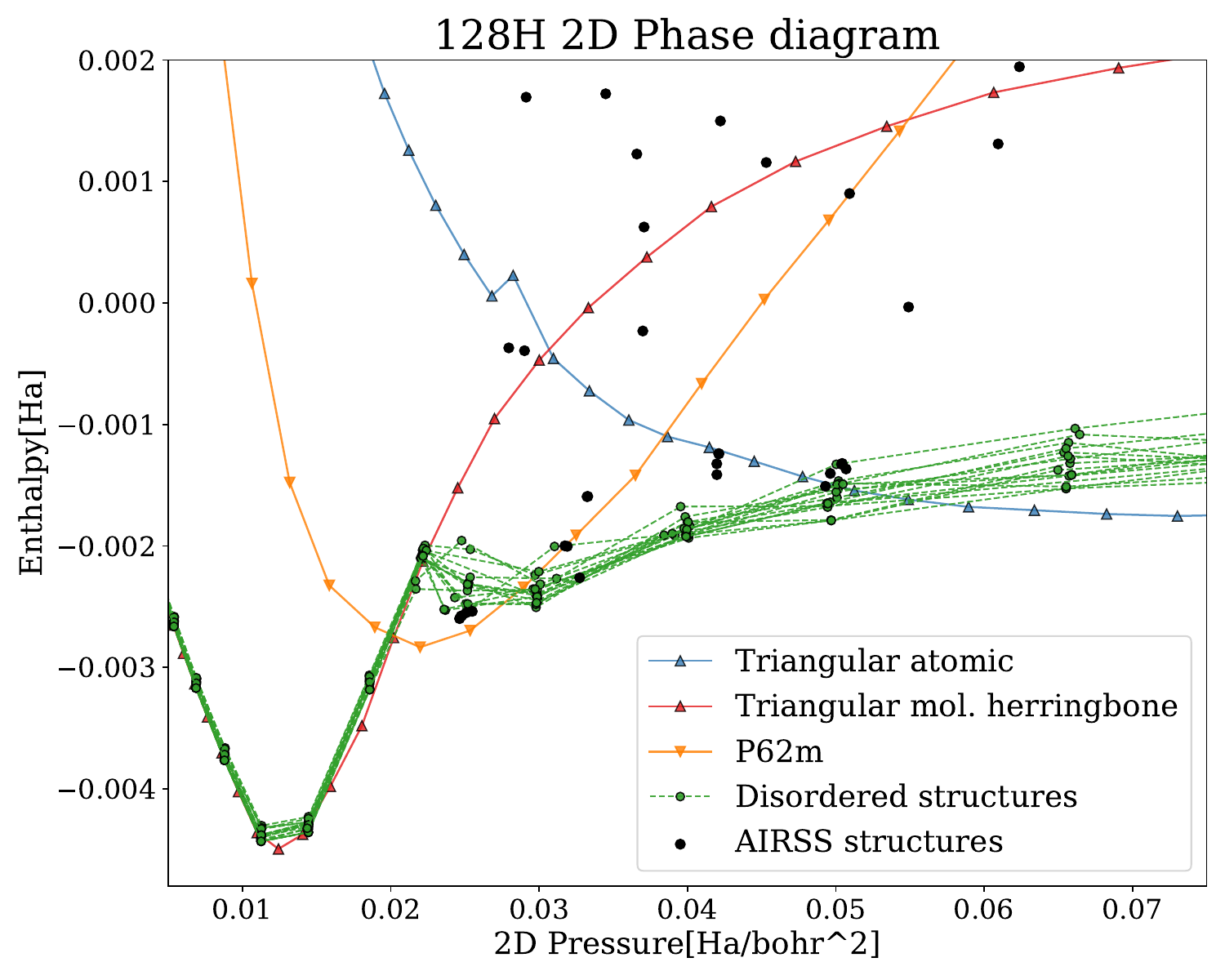}
    \hfill
    \includegraphics[width=0.49\linewidth]{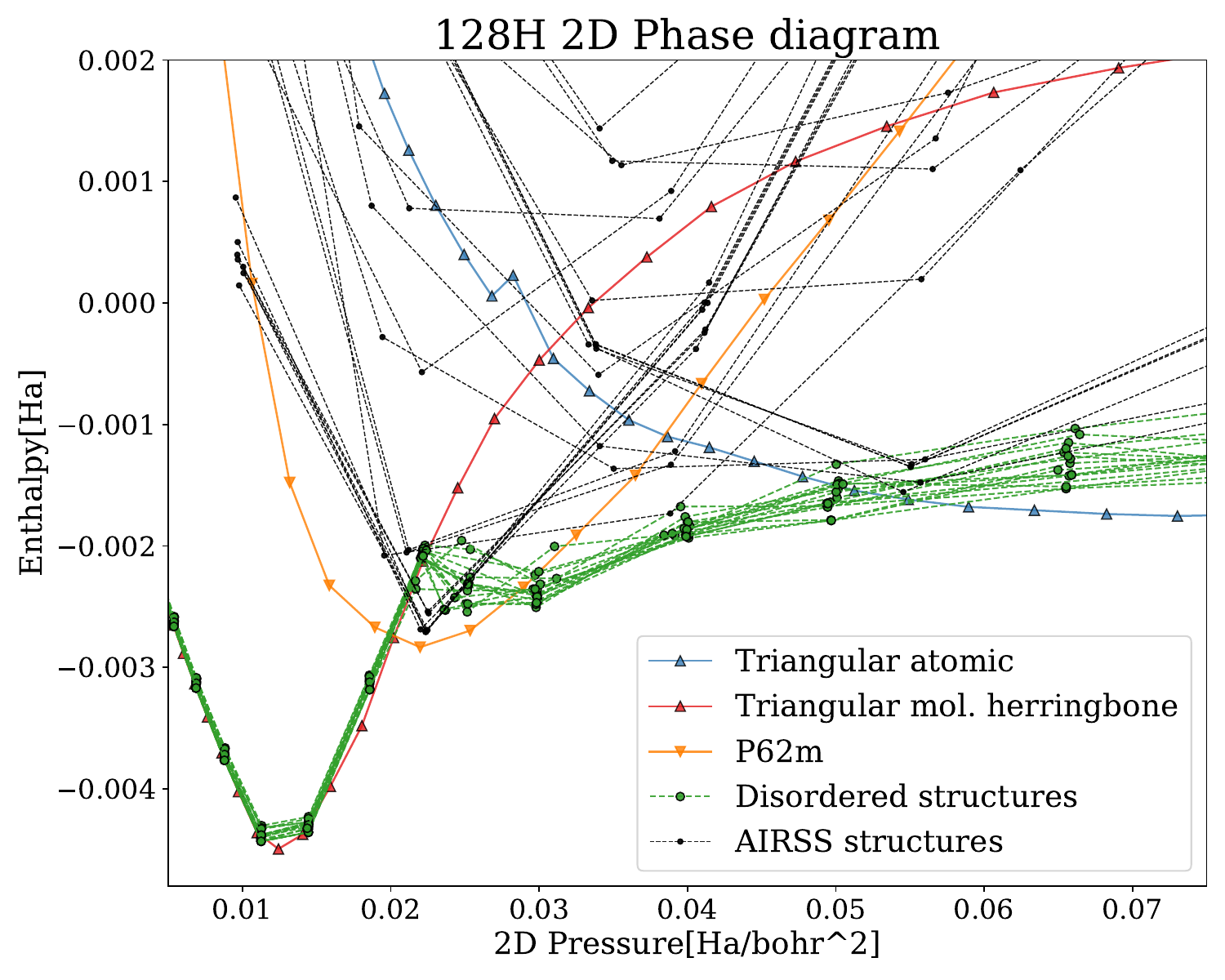}
    \caption{Enthalpy of candidate atomic structures generated by AIRSS compared to the phase diagram of 128H. Left: AIRSS generated structures (6-24 H atoms). Right: Each curve corresponds to a candidate AIRSS structure replicated to form a 128-atom supercell and evaluated across the pressure range. Notice that the green (disordered) data series overlap with the ``herringbone'' points below 0.02 Ha/Bohr$^{2}$ as the random initial structure indeed crystallize spontaneously in this molecular phase.}
    \label{fig:airss}
\end{figure}

We explored systems containing between 6 and 24 atoms per simulation cell, generating in total more than $10^4$ candidate structures across the investigated pressure range. The lowest-enthalpy candidates are shown in Fig.~\ref{fig:airss} (right).

For the bests candidate structures found by AIRSS, we also constructed a 128-atom supercell in order to evaluate its stability across the phase diagram. As shown in Fig.~\ref{fig:airss} (left), these ordered candidates do not outperform the disordered structures identified in our previous searches, except in a narrow region close to the boundary with the triangular molecular herringbone phase.

\clearpage

\section{Metal-Insulator Transition}
\begin{figure}[h]
    \centering
    \includegraphics[width=0.5\linewidth]{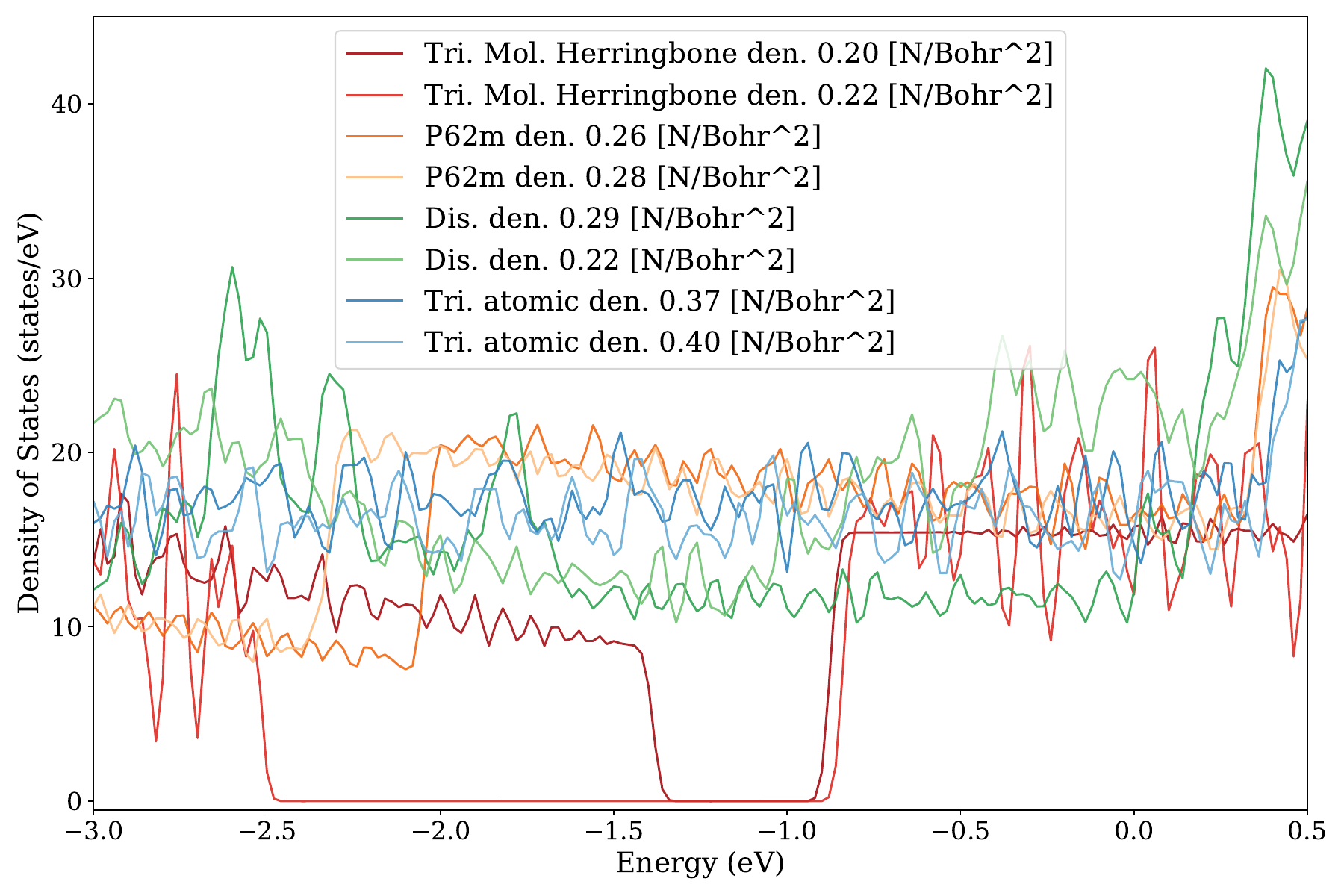}
    \caption{Density of States at increasing densities across the phase diagram of the favored lattice at that condition.}
    \label{fig:DOS}
\end{figure}

We investigate the metallic properties of the phases found.
To do so we calculate the density of state ( Fig.~\ref{fig:DOS}). The system is an insulator only for the Triangular Molecular Herringbone lattice, that is also the only molecular lattice. 
According to this analysis the insulator-to-metal transition coincides with the molecular-to-atomic transition.
Note that PBE is an xc-functional known to underestimate the bandgap.

\section{Complete results}
\begin{figure}
\centering
\includegraphics[width=0.9\linewidth]{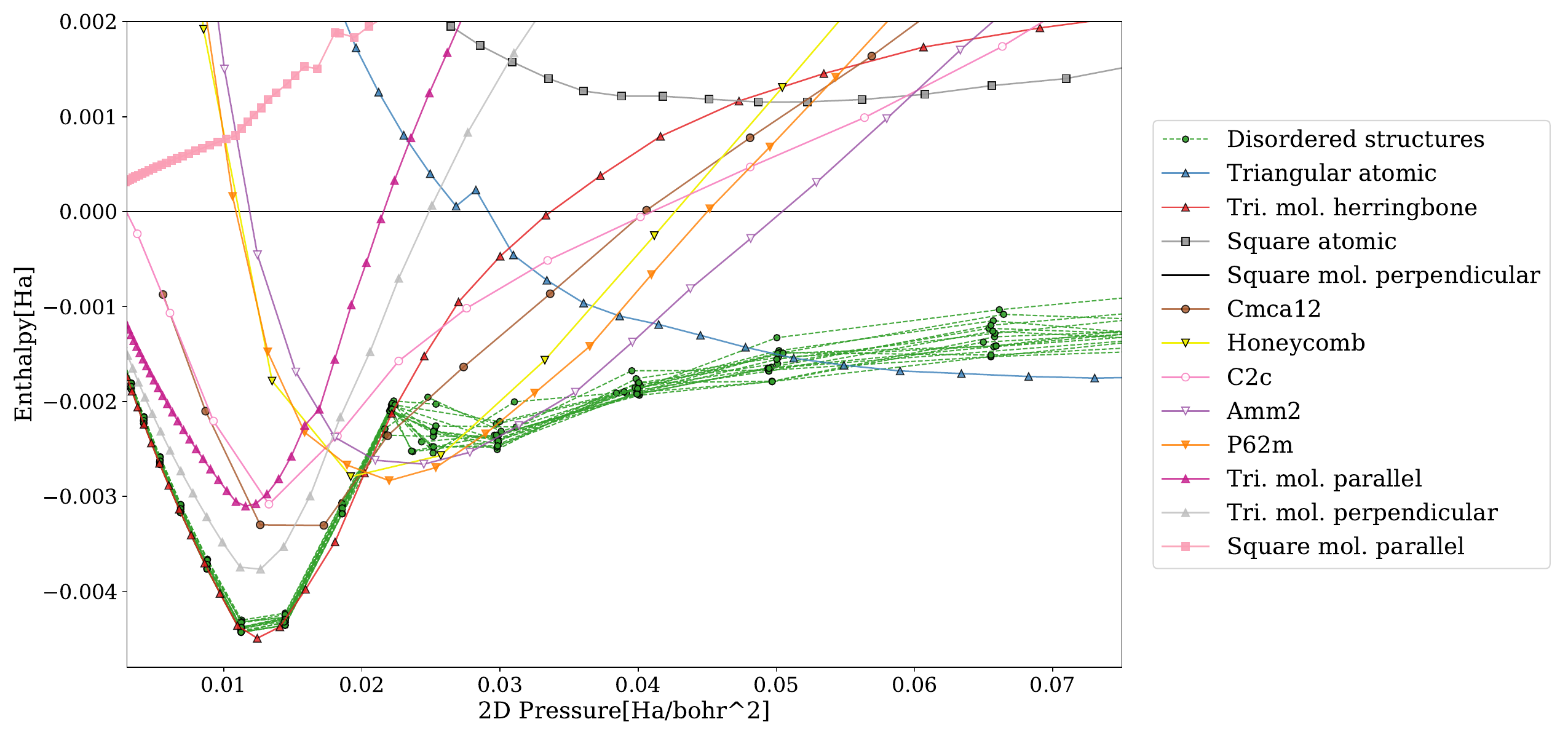}
\caption{Complete enthalpy phase diagram of 2D hydrogen for the 128-atom system, including all candidate structures considered in this work.}
\label{fig:complete_enthalpy}
\end{figure}

\begin{figure}
\centering
\includegraphics[width=0.9\linewidth]{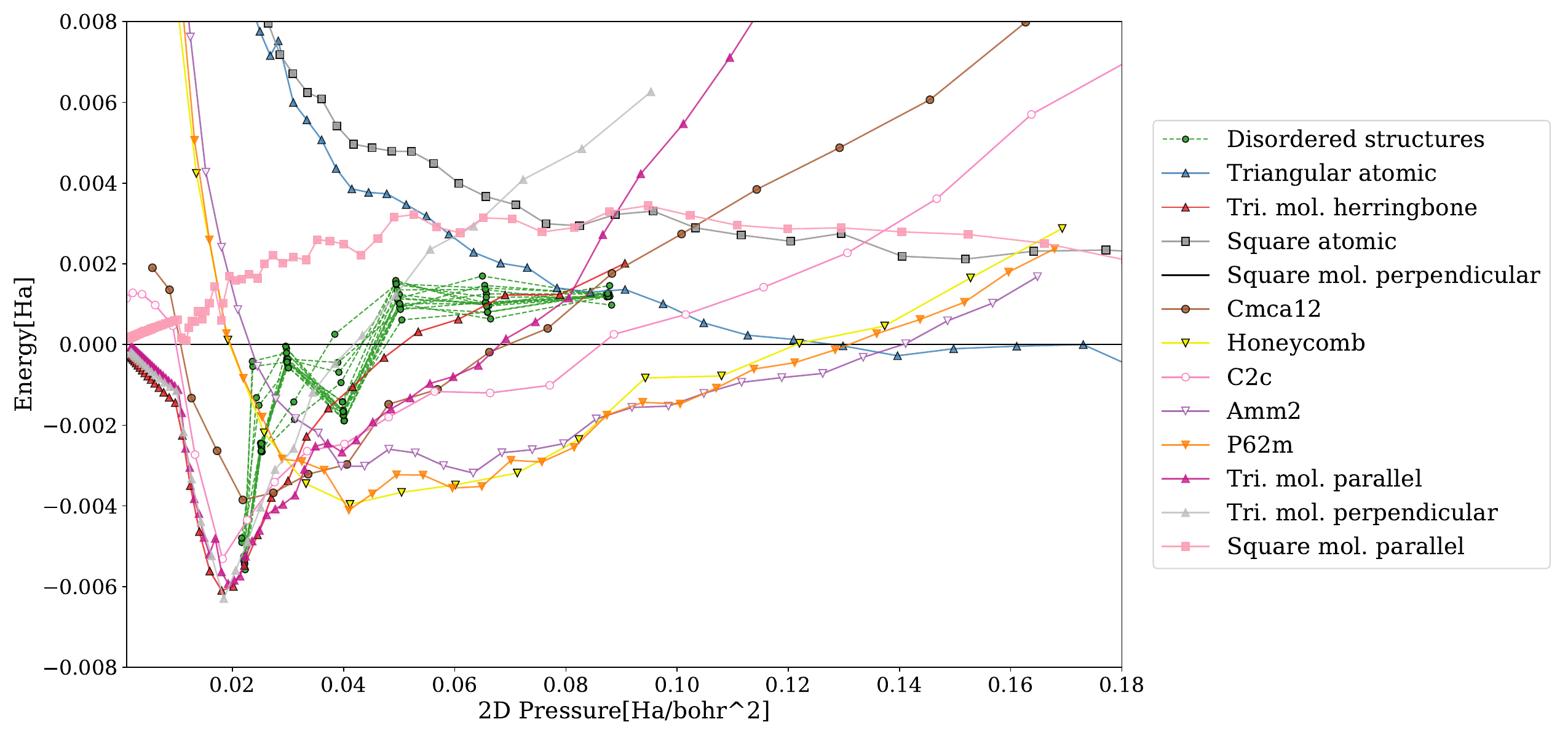}
\caption{Complete internal-energy phase diagram of 2D hydrogen for the 128-atom system, including all candidate structures considered in this work.}
\label{fig:complete_energy}
\end{figure}

Figures~\ref{fig:complete_enthalpy} and \ref{fig:complete_energy} present the complete set of structures considered in this work for the 128-atom system. Compared to Fig.~\ref{fig:phase_diagram} in the main text, these phase diagrams include additional candidate lattices that were omitted there for clarity.

Figure~\ref{fig:complete_enthalpy} shows the phase diagram constructed from the enthalpy, which is the relevant thermodynamic potential under pressure. The results confirm the stability of the triangular molecular herringbone phase at low pressure and the triangular atomic phase at high pressure, with the disordered structures becoming energetically favorable in the intermediate pressure regime.

For comparison, Fig.~\ref{fig:complete_energy} reports the corresponding phase diagram obtained using the internal energy alone. The resulting picture differs substantially from the enthalpy-based phase diagram. While the triangular molecular herringbone and triangular atomic phases remain the lowest energy structures at low and high pressure, respectively, the intermediate pressure region is instead dominated by honeycomb-like arrangements. In particular, the Honeycomb, Amm2, and P62m structures, which can all be viewed as distorted honeycomb lattices, exhibit lower internal energies than the disordered structures over a broad pressure range.

The preference for honeycomb-derived structures in terms of internal energy can be understood from their local bonding environment. In these lattices, each atom has three nearest neighbors. However, the reduced coordination number also results in a lower packing efficiency than in triangular lattices, where each atom has six nearest neighbors. Consequently, honeycomb-based structures occupy a larger area per atom and become increasingly penalized by the contribution to the enthalpy at finite pressure.

This comparison highlights the crucial role of the pressure term in determining thermodynamic stability. Although several ordered honeycomb-derived structures are energetically competitive and often exhibit lower internal energies than the disordered phase, their larger equilibrium area ultimately makes them less favorable in enthalpy, leading to the stabilization of the disordered phase in the intermediate pressure region.\\ 

\begin{figure}[h!]
\centering
\includegraphics[width=0.9\linewidth]{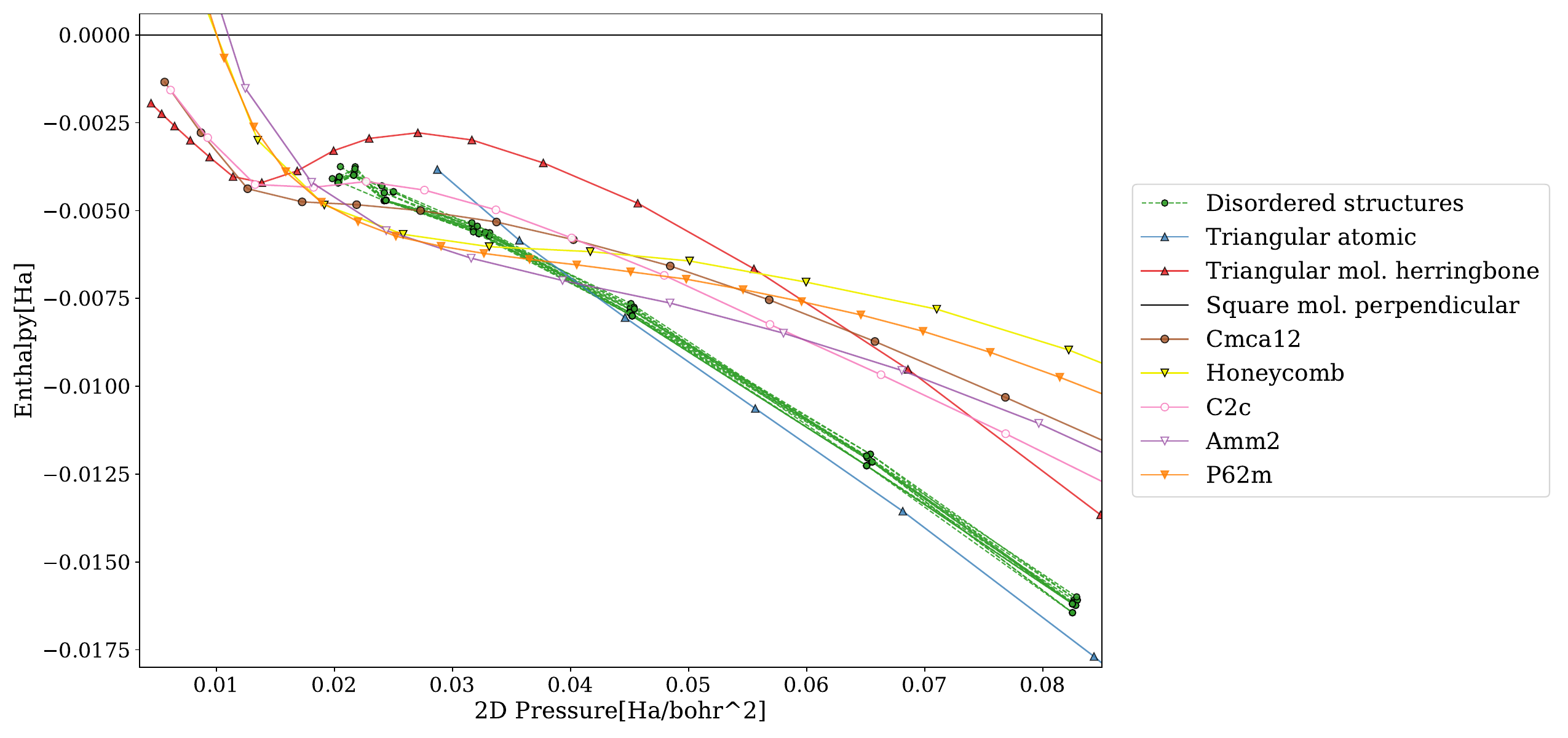}
\caption{Phase diagram of 2D hydrogen with 48 atoms.}
\label{fig:48H_phase_diag}
\end{figure}

\textbf{Small supercell phase diagram.}
In Fig.~\ref{fig:48H_phase_diag}, we show the phase diagram of 2DH obtained using 48 atoms. It is clear that the use of smaller supercells leads to a qualitatively different phase diagram, with the disordered phase being completely suppressed and an additional molecular phase, the $Cmca12$ lattice, appearing between the herringbone and $P62m$ phases.

\section{Details and results of the PIMD Simulations}

\subsection{PIMD Setup}
To perform path-integral molecular dynamics (PIMD) simulations of strictly two-dimensional hydrogen, we modified the NVT integrator of \textsf{i-PI}. Specifically, the out-of-plane ($z$) components of both the atomic displacements and momenta were constrained to zero at each integration step, thereby restricting the dynamics to the $xy$ plane.

To determine the number of beads required for converged quantum statistics, we considered an isolated H$_2$ molecule at T=100 K. This system provides a convenient benchmark because the radial distribution function $g(r)$ can be obtained analytically within the harmonic approximation.

The analytical radial distribution functions, $g(r)$, for the internuclear bond length of an isolated H$_2$ molecule are derived from the thermal density matrix of a one-dimensional harmonic oscillator. To accurately model the system, the ground-state Potential Energy Surface (PES) of the H$_2$ molecule was computed at the PBE level of theory and fitted to a harmonic potential. This fitting procedure yielded an equilibrium internuclear distance of $r_0 = 1.4284$ Bohr, a harmonic force constant of $k = 0.3419$ a.u., and a base reference frequency of $\omega_0 = 0.5847$ a.u. Within the theoretical framework, $r$ represents the computed Euclidean internuclear distance, $\mu$ is the reduced mass of the H$_2$ molecule defined as $\mu = m_H / 2$, and $\omega$ is the fundamental physical vibrational frequency given by $\omega = \sqrt{k/\mu}$. The thermal effects are governed by the inverse thermodynamic temperature $\beta = 1 / (k_B T)$, and $\mathcal{N}$ denotes the respective normalization constants required to ensure the total probability integrates to unity.

The fundamental difference between the one-dimensional probability density and the true 2D or 3D radial distributions arises from the Jacobian of transformation from Cartesian to polar or spherical coordinates. 
 
The classical and quantum radial distribution functions in 3D are given by:
\begin{equation}
    g_{\text{cl}}^{\text{3D}}(r) = \mathcal{N}_{\text{cl}}^{\text{3D}} \, r^2 \exp\left[ -\frac{\beta k}{2} (r - r_0)^2 \right]
\end{equation}
\begin{equation}
    g_{\text{qm}}^{\text{3D}}(r) = \mathcal{N}_{\text{qm}}^{\text{3D}} \, r^2 \exp\left[ -\mu \omega \tanh\left(\frac{\beta \omega}{2}\right) (r - r_0)^2 \right]
\end{equation}

Instead, when the molecule is constrained to rotate within a two-dimensional plane, the geometric volume is proportional to the circumference of a circle, reducing the radial phase-space factor to $r$. The corresponding classical and quantum distributions in 2D are:
\begin{equation}
    g_{\text{cl}}^{\text{2D}}(r) = \mathcal{N}_{\text{cl}}^{\text{2D}} \, r \exp\left[ -\frac{\beta k}{2} (r - r_0)^2 \right]
\end{equation}
\begin{equation}
    g_{\text{qm}}^{\text{2D}}(r) = \mathcal{N}_{\text{qm}}^{\text{2D}} \, r \exp\left[ -\mu \omega \tanh\left(\frac{\beta \omega}{2}\right) (r - r_0)^2 \right]
\end{equation}

\begin{figure}
    \centering
    \includegraphics[width=0.49\linewidth]{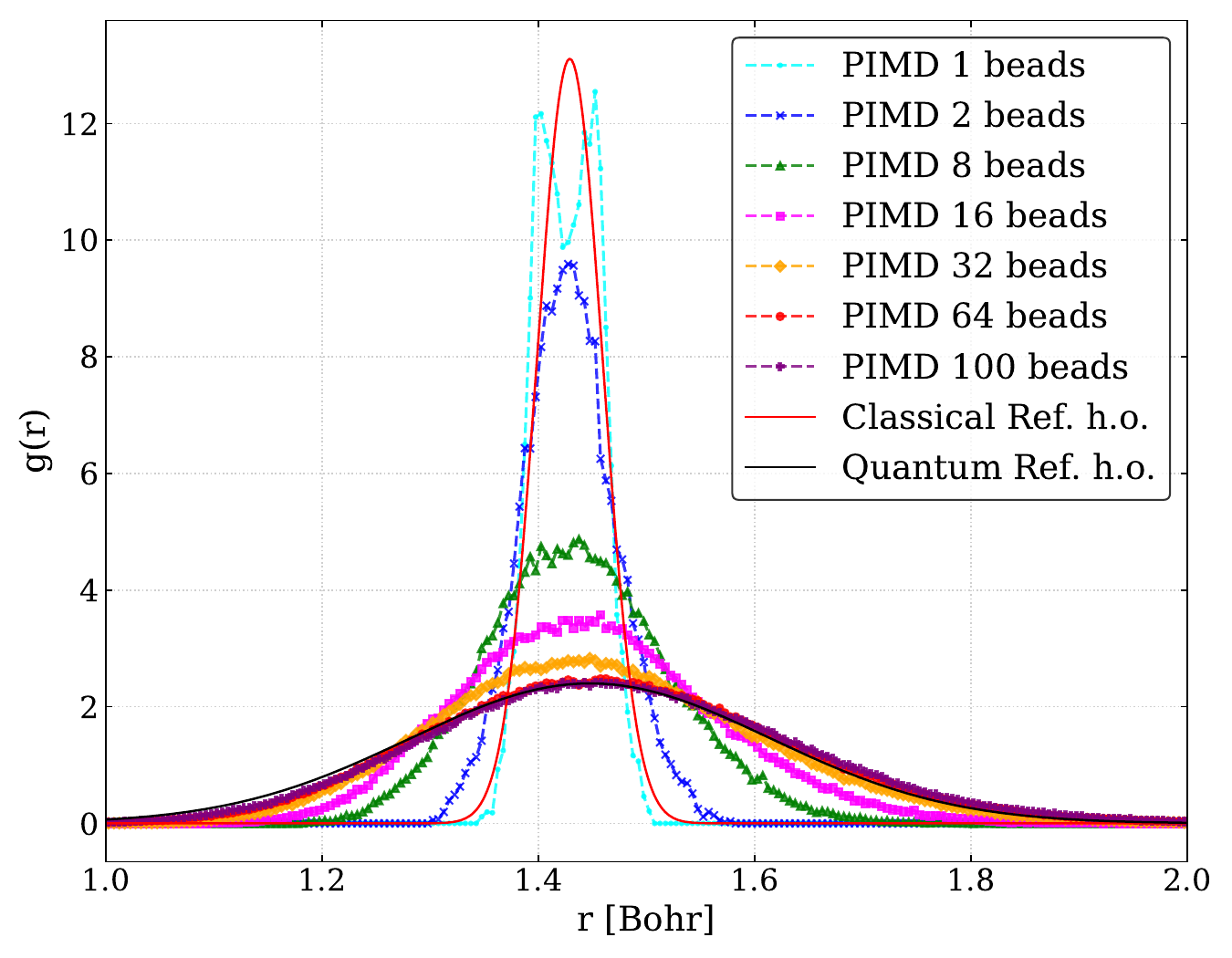}
    \hfill
    \includegraphics[width=0.49\linewidth]{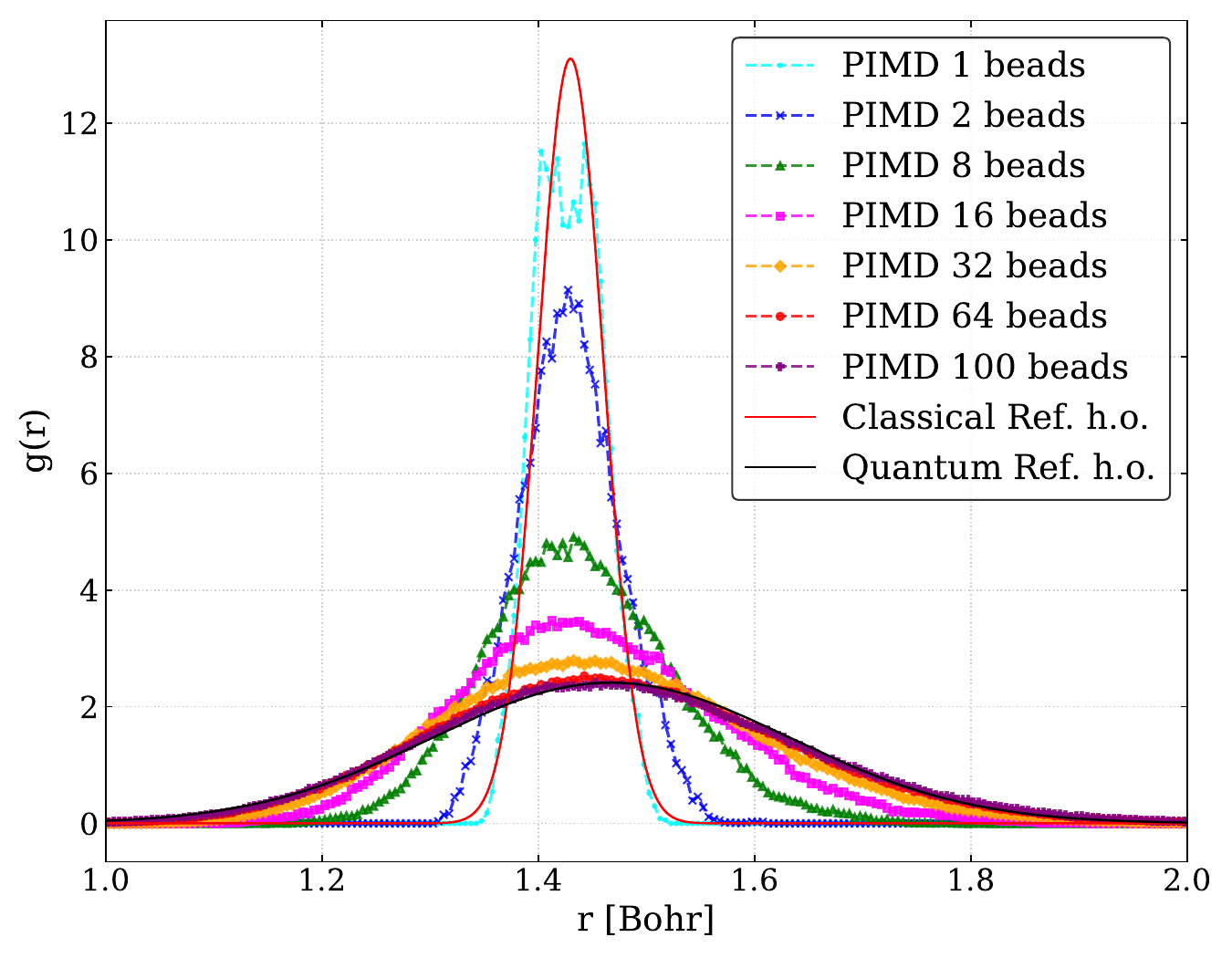}
    \caption{Convergence of the radial distribution function of an isolated H$_2$ molecule at 100 K with respect to the number of PIMD beads. \emph{Left:} two-dimensional case. \emph{Right:} three-dimensional case. The analytical quantum distribution is accurately reproduced with 64 beads in both cases.}
    \label{fig:H2_gr}
\end{figure}

Figure~\ref{fig:H2_gr} compares the analytical quantum distributions with PIMD simulations performed using different numbers of beads. In both the 2D and 3D cases, the radial distribution function is converged at 64 beads, which was therefore adopted in all production simulations.

\subsection{PIMD Results}

We perform NVT PIMD simulations starting from a 120-atoms $P62m$ configuration and 128-atoms disordered configuration. The density is set to be 0.29 atoms$/Bohr^2$ for both configurations, that leads to an equilibrated 2D pressure of 0.0354 $Ha/Bohr^2$ for the $P62m$ configuration and 0.0346 $Ha/Bohr^2$ for the disordered configuration. We observe that the $P62m$ system evolves away from this lattice structure reaching the same energy and structure factor of the disordered system.

\begin{figure}
    \centering
    \includegraphics[width=0.8\linewidth]{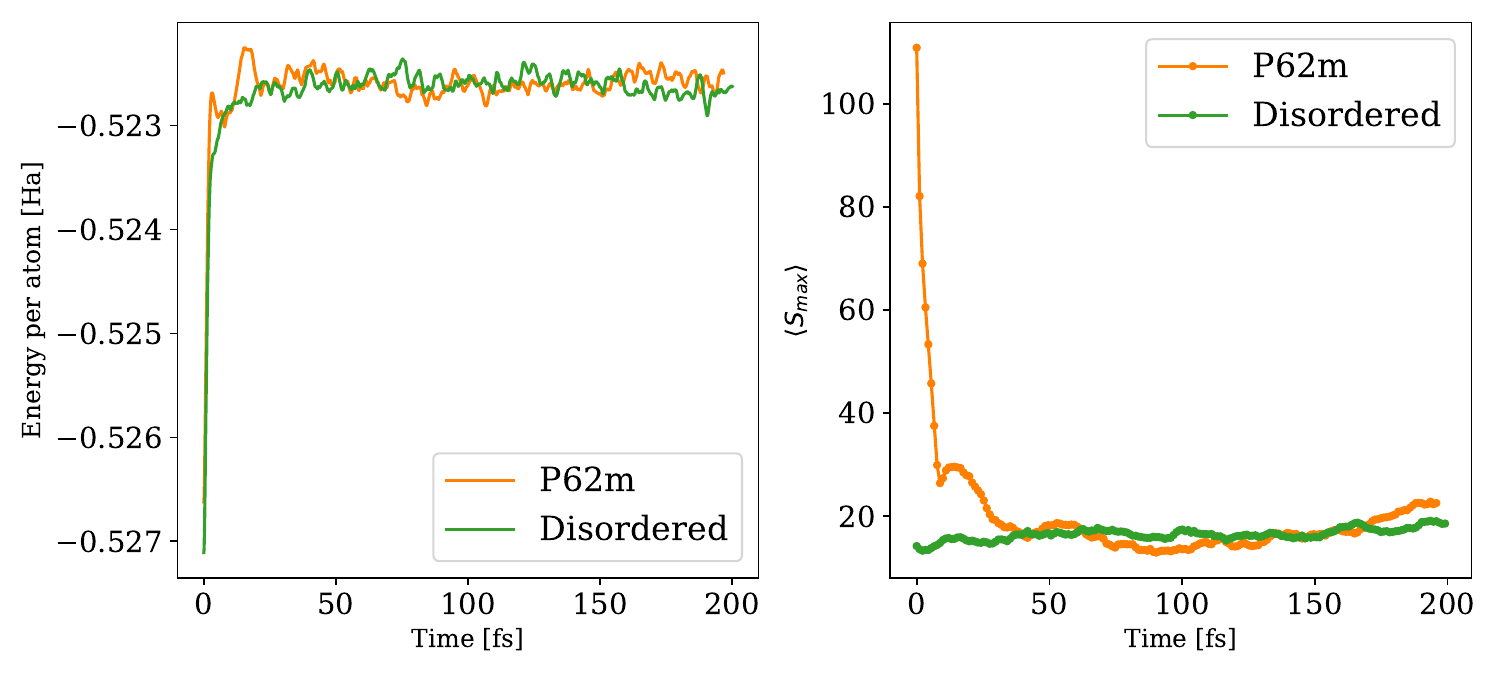}
    \caption{2D PIMD at 100K starting from $P62m$ atomic lattice (orange) and a disordered structure (green). After a small equilibration time the system find the same equilibrium condition in a disordered state. \emph{Left:} Energy per atom over time. \emph{Right:} Maximum of the Structure factor over time. }
    \label{fig:pimd}
\end{figure}

\begin{figure}
    \centering
    \includegraphics[width=0.8\linewidth]{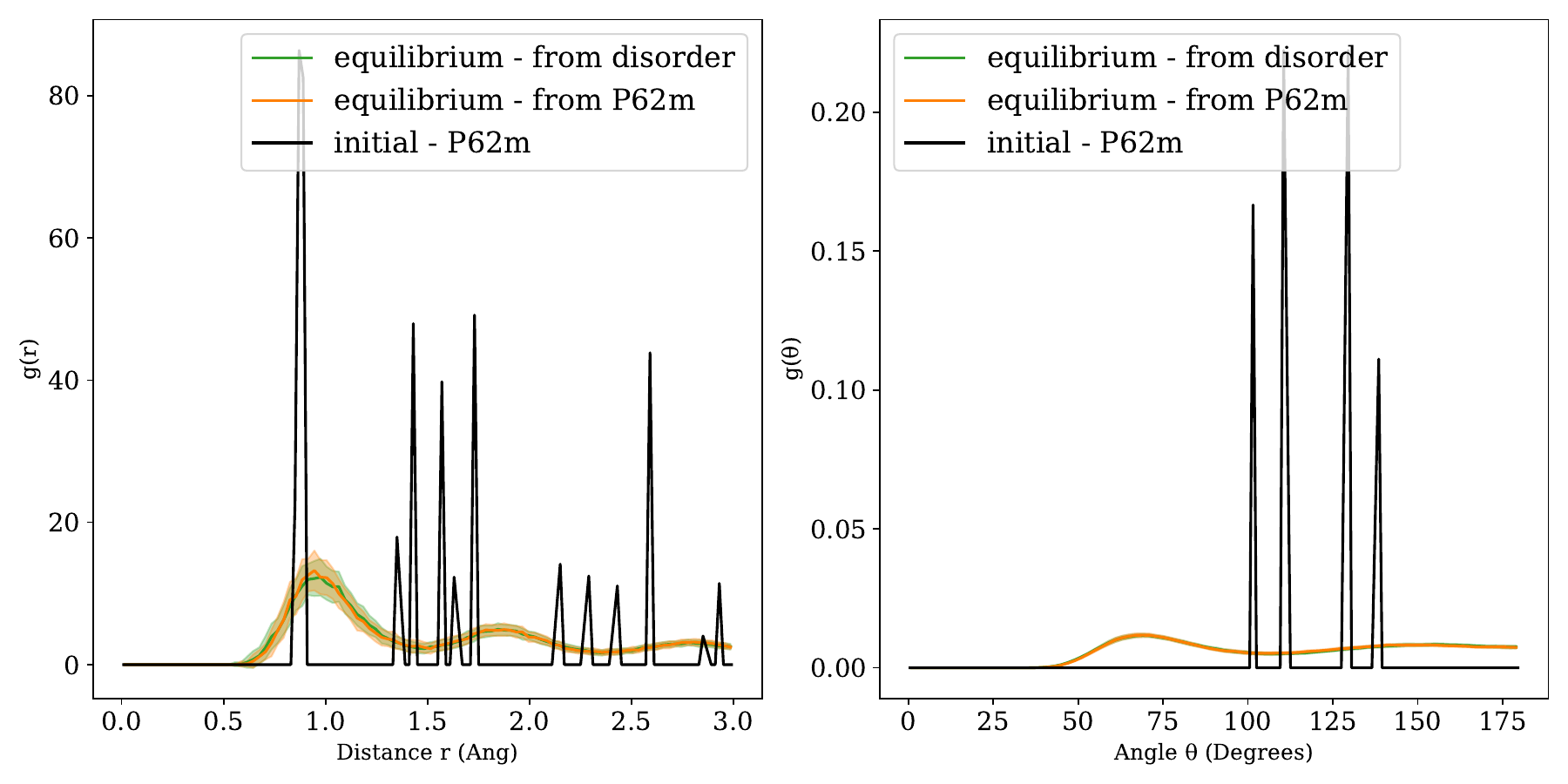}
    \caption{Radial distribution function and Angular distribution function of the PIMD. The black line represents the initial crystal configuration, the orange and green line represents the same quantity averaged over the last 500 steps for all the replicas of the PIMD starting from $P62m$ and from a disordered configuration, respectively. The shaded region correspond to one sigma standard deviation over the 64 replicas of the PIMD. To compute  $g(\theta)$ , we calculated the angles formed between each atom and all possible pairs of its neighbors located within a first-coordination-shell cutoff of 1 Ang. The angular distribution functions at equilibrium starting from a crystalline $P62m$ lattice or from a disordered structure are perfectly in agreement.  }
    \label{fig:pimd_distrib}
\end{figure}

\begin{figure}
    \centering
    \includegraphics[width=0.49\linewidth]{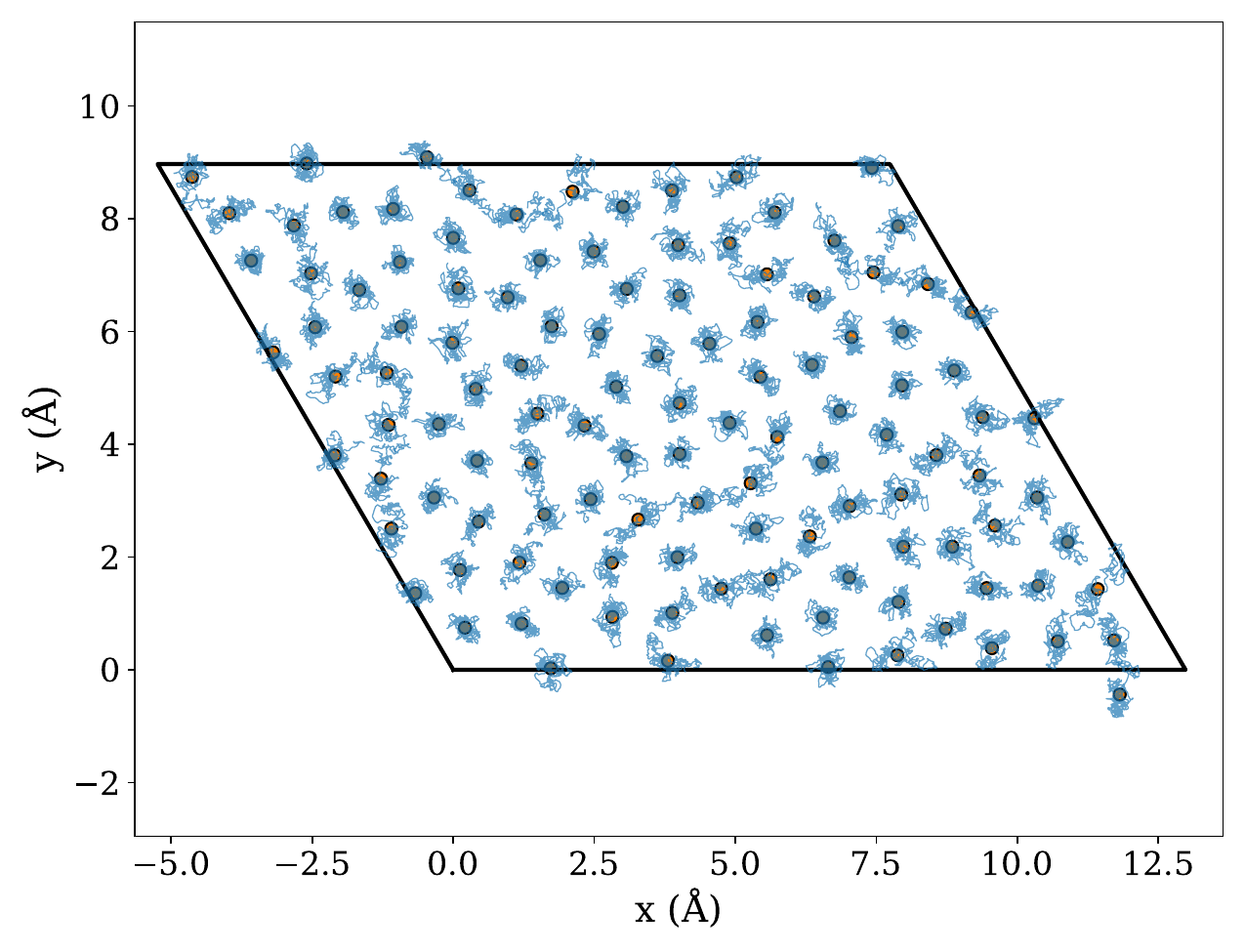}
    \hfill
    \includegraphics[width=0.49\linewidth]{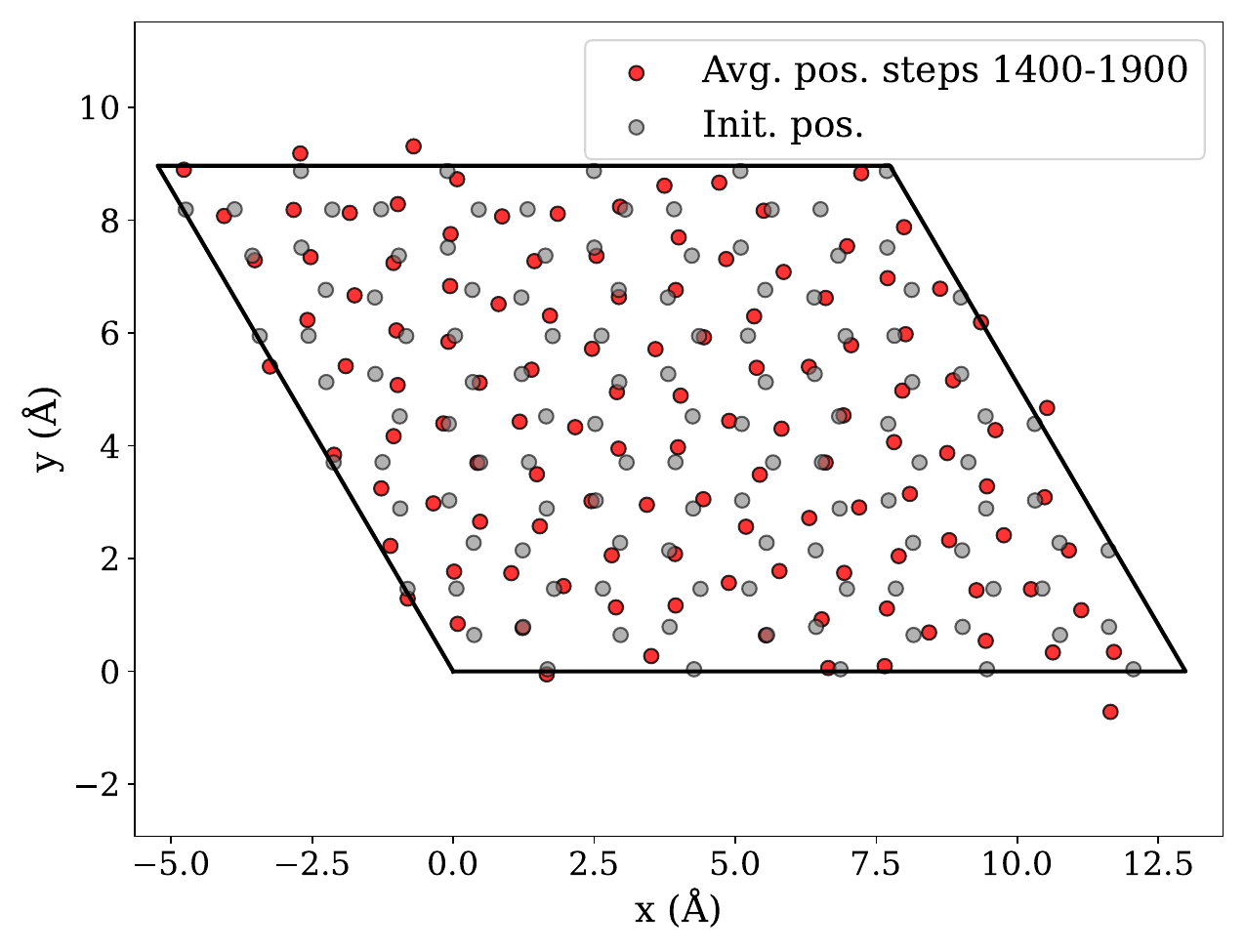}
    \caption{\emph{(Left): }Average position over the last 500 steps of the $P62m$ PIMD. Orange dots represents the average position of the atoms over the last 500 steps of the PIMD (here is shown only one replica). The blue lines represent the displacement of the atoms over the full PIMD (here again is shown the same replica of the average position). (\emph{Right}): Average atomic positions during PIMD across different phase windows.} 
    \label{fig:pimd_atoms}
\end{figure}

The dynamical structure factor decreases rapidly (Fig~\ref{fig:pimd}), while the broadened peaks of the equilibrated pair correlation functions ($g(r)$) and ($g(\theta)$) are not consistent with the initial structure (Fig.~\ref{fig:pimd_distrib}).
Both simulations equilibrates to the same distributions.
The trajectories in the $x-y$ plane also suggest that the atoms are not simply undergoing fluctuations around the $P62m$ crystal lattice (Fig.~\ref{fig:pimd_atoms}).

For the PIMD we have set the DFT energy cut-off to 120.0 Ry and the k-point sampling grid to $4\times4\times1$. These parameters allows us to save computational time over the very expensive 64 beads PIMD, while still being highly accurate (see above).

\end{document}